\documentclass[12pt]{article}
\usepackage{amssymb,amsmath,epsfig}

\begin{document}

\title{\bf Energy Constraints and $F(T,T_{G})$ Cosmology}
\author{Saira Waheed \thanks{smathematics@hotmail.com} and M.
Zubair $^{a}$
\thanks{mzubairkk@gmail.com; drmzubair@ciitlahore.du.pk}\\
Department of Mathematics, University of the Punjab,\\
Quaid-e-Azam Campus, Lahore-54590, Pakistan. \\
$^{a\dag}$Department of Mathematics, COMSATS Institute of
\\Information Technology Lahore, Pakistan}

\date{}

\maketitle
\begin{abstract}
The present paper is elaborated to discuss the energy condition
bounds in a modified teleparallel gravity namely $F(T,T_{G})$,
involving torsion invariant $T$ and contribution from a term $T_G$,
the teleparallel equivalent of the Gauss-Bonnet term. For this
purpose, we consider flat FRW universe with matter contents as
perfect fluid. We formulate the SEC, NEC, WEC and DEC in terms of
some cosmic parameters including Hubble, deceleration, jerk and snap
parameters. By taking two interesting models for $F(T,T_{G})$ and
some recent limits of these cosmic parameters, we explore the
constraints on the free parameters present in both assumed models.
We also discuss these constraints graphically in terms of cosmic
time by taking power law cosmology into account.
\end{abstract}

\section{Introduction}

One of the most revolutionary investigation of the previous century
is the accelerated expanding behavior of cosmos that motivates the
scientists in a new direction. This interesting fact is
substantiated by numerous observational probes \cite{1}-\cite{7}
etc. and leads to the existence of a new cryptic dominant ingredient
in the cosmic matter distribution refereed as dark energy (DE). In
order to comprehend the nature of this new sort of energy, numerous
attempts are made by incorporating some new terms in the
Einstein-Hilbert Lagrangian density either in the matter sector or
the gravitational sector of the action. As a result, these
techniques offered a huge group of DE models including Chaplygin gas
\cite{8}, cosmological constant \cite{9}, tachyon fields \cite{10},
quintessence \cite{11}, k-essence \cite{12}, modified theories like
$f(R)$ gravity \cite{13}, Gauss-Bonnet gravity \cite{14}, $f(T)$
theory \cite{15}, $f(R,T)$ gravity \cite{16} and scalar-tensor
theories \cite{17} that have numerous distinct and interesting
cosmological applications.

Another interesting modification of Einstein's relativity is
obtained by introducing torsional formulation (torsion scalar $T$
which is obtained by contraction of torsion tensor) to explain the
gravitational effects instead of curvature scalar \cite{18}. In such
a gravitational framework, Lagrangian density includes curvature
less Weitzenb$\ddot{o}$ck connection as a replacement of torsion
less Levi-Civita connection. This theory is referred as TEGR
(teleparallel equivalent of general relativity) and has widespread
applications in cosmology. Another comprehensive form of this theory
has been proposed in literature \cite{19} by replacing torsion
scalar with a general function $f(T)$, known as $f(T)$ theory of
gravity. Numerous significant cosmological aspects of this theory
has been discussed in literature \cite{19,20}. Its another useful
version is obtained by considering higher-torsion corrections just
like the case of Gauss-Bonnet term \cite{21} (arising from
higher-curvature corrections), Lovelock combinations \cite{22}, Weyl
combinations \cite{23}. Based on this concept, Kofinas and Saridakis
\cite{24,25} proposed a novel theory namely $F(T_{G})$ gravity and
then its generalized form $F(T, T_{G})$ gravity and they also
discussed its cosmological significance.

Energy condition bounds are used to explore the constraints on the
free parameters arising from different DE models. These constraints
has been discussed in various contexts like $f(R)$ gravity
\cite{26}, $f(T)$ theory \cite{27}, $f(G)$ theory \cite{27*}, $f(R)$
gravity with nonminimal interaction with matter \cite{28},
$f(R,\mathcal{L}_m)$ gravity \cite{29} and Brans-Dicke theory
\cite{30}. In this regard, Sharif and Saira \cite{31} have discussed
the energy conditions in the most general scalar-tensor gravity with
dynamical equations involving second-order derivatives of scalar
field for perfect fluid FRW geometry. Sharif and Zubair \cite{32}
have explored these constraints in a general theory $f(R,T)$ theory
involving the trace of energy-momentum tensor. They also discussed
the stability criteria for such configuration using power law
cosmology. They have also examined some models using energy
inequalities for $f(R,T,R_{\mu\nu}T^{\mu\nu})$ gravity \cite{33}.
Recently, we have examined the energy bounds in a modified theory
based on the non-minimal interaction of torsion scalar and perfect
fluid matter using power law form of FRW cosmology \cite{34}. We
have derived the general inequalities involving different cosmic
parameters and discussed them graphically.

In the present work, we are interested to discuss the energy
constraints in $F(T, T_{G})$ gravity using FRW universe model filled
with perfect fluid matter. We derive these constraints in terms of
cosmic parameters for two different proposed models of $F(T,
T_{G})$. The paper is designed in this layout. In the next section,
we provide a general introduction of $F(T, T_{G})$ theory and
discuss the basic formulation of energy bounds. Section \textbf{3}
is devoted to study these constraints for two different models of
$F(T, T_{G})$. Here we provide the graphical illustration of the
obtained inequalities. Finally, we summarize the whole discussion.

\section{Introduction to $F(T,T_{G})$ Cosmology and General Formulation of Energy Bounds}

In this section, we briefly explain some basic ingredients of TEGR
and hence of $F(T,T_{G})$. Here we also discuss the basic
formulation of energy constraints. In tangent components, torsion
and curvature tensor are defined as
\begin{eqnarray}\label{1}
T^{a}_{bc}&=&\omega^{a}_{cb}-\omega^{a}_{bc}-C^{a}_{bc},\\\label{2}
R^{a}_{bcd}&=&\omega^{a}_{bd,c}-\omega^{a}_{bc,d}+\omega^{e}_{bd}\omega^{a}_{ec}
-\omega^{e}_{bc}\omega^{a}_{ed}-C^{e}_{cd}\omega^{a}_{be},
\end{eqnarray}
where the source of parallel transportation, connection 1-form
$\omega^a_b(x^\mu)$ in terms of vielbein field is given by
$\omega^{a}_{b}=\omega^{a}_{b\mu}dx^\mu=\omega^{a}_{bc}e^{c},$ while
$C^{c}_{ab}=e_{a}^{\mu}e_{b}^{\nu}(e^{c}_{\mu,\nu}-e^{c}_{\nu,\mu})$
denote the structure coefficients arising from the veilbein
commutation defined by
\begin{equation*}
[e_{a}, e_{b}]=C^{c}_{ab}e_{c}.
\end{equation*}
Also, $g_{\mu\nu}=\eta_{ab}e^{a}_{\mu}e^{b}_{\nu},$ where
$\eta_{ab}$ is the Minkowski metric. The contorsion tensor is
defined in terms of torsion tensor as follows
\begin{eqnarray}\label{3}
\mathcal{K}_{abc}=\frac{1}{2}(T_{cab}-T_{bca}-T_{abc})=-\mathcal{K}_{bac}.
\end{eqnarray}
In order to be consistent with the condition $R^{a}_{bcd}=0$
(teleparallelism condition), we express the Weitzenb$\ddot{o}$ck
connection as follows
$$\tilde{\omega}^{\lambda}_{\mu\nu}=e_{a}^{\lambda}e^{a}_{\mu,\nu},$$
while in terms of Levi-Civita connection, the Ricci scalar $R$ is
given by
\begin{eqnarray}\nonumber
e\bar{R}=-eT+2(eT_{\nu}^{\nu\mu})_{,\mu},
\end{eqnarray}
where
$$e=det(e^{a}_{\mu})=\sqrt{|g|},\quad T=\frac{1}{4}T^{\mu\nu\lambda}_{\mu\nu\lambda}
+\frac{1}{2}T^{\mu\nu\lambda}T_{\lambda\nu\mu}-T_{\nu}^{\nu\mu}T^{\lambda}_{\lambda\mu}.$$
Consequently, the Lagrangian density describing TEGR in D-dimensions
is given by
\begin{eqnarray}\label{4}
S_{tel}=-\frac{1}{2\kappa_{D}^2}\int_M d^Dx eT.
\end{eqnarray}
In a recent paper \cite{24}, teleparallel equivalent of Gauss-Bonnet
theory has been proposed involving a new torsion scalar $T_{G}$,
where, in Levi-Civita connection, the Gauss-Bonnet term is defined
by
\begin{equation}
e\bar{G}=eT_{G}+\verb"total diverg"
\end{equation}
and the corresponding action takes the following form
\begin{eqnarray}\label{5}
S_{tel}=-\frac{1}{2\kappa_{D}^2}\int_M d^Dx eT_{G}.
\end{eqnarray}
Since both theories $f(T)$ and $f(T_G)$ arise independently,
therefore a comprehensive theory involving both $T$ and $T_G$ as
basic ingredient has been proposed by Kofinas and Saridakis defined
by the following action
\begin{eqnarray}\label{6}
S_{tel}=-\frac{1}{2\kappa_{D}^2}\int_M d^Dx eF(T,T_{G}).
\end{eqnarray}
In some certain limits of the function $F(T,T_{G})$, other theories
like GR, TEGR, Einstein-Gauss-Bonnet theory etc. can be discussed.

Energy constraints have many useful applications in GR as well as in
modified gravity theories (discussion of various cosmological
geometries). These inequalities are firstly formulated in the
context of GR for the derivation of some general results involving
strong gravitational fields. In GR, four types of energy constraints
are formulated using a well-known geometrical results refereed as
Raychaudhuri equation (explaining the dynamics of matter bits).
These constraints are labeled as WEC, DEC, NEC and SEC. In a
spacetime manifold, Raychaudhuri equation provides the temporal
evolution of expansion scalar as a linear combination of Ricci
tensor $R_\mu\nu$, shear tensor $\sigma^{\mu\nu}$ and rotation
$\omega^{\mu\nu}$ given by
\begin{eqnarray}\label{7}
\frac{d\theta}{d\tau}=-\frac{1}{3}\theta^2-\sigma_{\mu\nu}\sigma^{\mu\nu}
+\omega_{\mu\nu}\omega^{\mu\nu}-R_{\mu\nu}u^\mu u^\nu,\\\label{8}
\frac{d\theta}{d\tau}=-\frac{1}{3}\theta^2-\sigma_{\mu\nu}\sigma^{\mu\nu}
+\omega_{\mu\nu}\omega^{\mu\nu}-R_{\mu\nu}k^\mu k^\nu,
\end{eqnarray}
where $u^\mu$ and $k^{\nu}$ denote the tangent vectors to timelike
and lightlike curves in the congruence. The attractive nature of
gravity makes the congruence geodesic convergent (congruence gets
closer to each other) and therefore leads to
$\frac{d\theta}{d\tau}<0$. Further, in some certain limits
\cite{35}, the quadratic terms can be ignored and hence leads to
inequalities
$$R_{\mu\nu}u^\mu u^\nu\geq0,\quad R_{\mu\nu}k^\mu k^\nu\geq0.$$
These inequalities can be further formulated in terms of
energy-momentum tensor and its trace by the inversion of the
gravitational field equations as follows
\begin{equation}\label{9}
(T_{\mu\nu}-\frac{T}{2}g_{\mu\nu})u^\mu u^\nu\geq0,\quad
(T_{\mu\nu}-\frac{T}{2}g_{\mu\nu})k^\mu k^\nu\geq0.
\end{equation}
For the ideal case, i.e., perfect fluid matter, these inequalities
provide the SEC, NEC, WEC, DEC as follows:
\begin{eqnarray}\nonumber
\textbf{NEC}:\quad&&\rho+P\geq0,\\\nonumber
\textbf{SEC}:\quad&&\rho+P\geq0,\quad\rho+3P\geq0,\\\nonumber
\textbf{WEC}:\quad&&\rho\geq0,\quad\rho+P\geq0,\\\label{10}
\textbf{DEC}:\quad&&\rho\geq0,\quad\rho\pm P\geq0.
\end{eqnarray}
In case of modified theories of gravity, these constraints can be
defined by simply replacing the $\rho$ by $\rho_{eff}$ and $P$ by
$P_{eff}$ with the assumption that the total matter contents behave
like perfect fluid.

\section{Energy Constraints in $F(T, T_G)$ Cosmology}

Here we first formulate the gravitational field equations
corresponding to the action (\ref{6}) for FRW geometry with perfect
fluid matter and then formulate the energy conditions for such a
configuration. In the presence of matter sector, the action
(\ref{6}) takes the following form
\begin{eqnarray}\label{11}
S_{tel}=-\frac{1}{2\kappa_{D}^2}\int_M d^Dx eF(T,T_{G})+S_m.
\end{eqnarray}
We consider the flat FRW universe model with $a(t)$ as expansion
radius given by
\begin{eqnarray}\label{12}
ds^2=-dt^2-a^2(t)(dx^2+dy^2+dz^2).
\end{eqnarray}
The diagonal vierbein and the dual vierbein for this metric are
\begin{eqnarray}\nonumber
e^a_\mu=diag(1, a(t), a(t), a(t)),\\\nonumber e^\mu_a=(1, a^{-1}(t),
a^{-1}(t), a^{-1}(t)),
\end{eqnarray}
while the corresponding determinant is given by $e=a(t)^3$. The
torsion scalar and Gauss-Bonnet equivalent term $T_G$ for this
geometry in terms of Hubble parameter $H=\frac{\dot{a}}{a}$ are
\begin{equation}\label{13}
T=6H^2, \quad T_G=24H^2(\dot{H}+H^2).
\end{equation}
The gravitational field equations corresponding to the action
(\ref{13}) for this geometry are given by
\begin{eqnarray}\label{14}
&&F-12H^2F_{T}-T_{G}F_{T_{G}}+24H^3\dot{F}_{T_{G}}=2\kappa^2\rho_m,\\\nonumber
&&F-4(\dot{H}+3H^2)F_T-4H\dot{F}_T-T_GF_{T_{G}}+\frac{2}{3H}T_G\dot{F}_{T_{G}}
+8H^2\ddot{F}_{T_{G}}=-2\kappa^2P_m,\\\label{15}
\end{eqnarray}
where $\rho_m$ and $P_m$ indicates the density and pressure of
ordinary matter, respectively described by the ideal case of
energy-momentum tensor given as follows
\begin{equation}\label{15*}
T_{{\mu}{\nu}}=({\rho_m}+P_m)u_{\mu}u_{\nu}-P_mg_{{\mu}{\nu}},
\end{equation}
and $\kappa$ is the gravitational coupling constant. The
gravitational field equations can also be written as
\begin{eqnarray}\label{16}
H^2&=&\frac{\kappa^2}{3}(\rho_m+\rho_{DE})\\\label{17}
\dot{H}&=&-\frac{\kappa^2}{2}(\rho_m+P_m+\rho_{DE}+P_{DE}),
\end{eqnarray}
where $\rho_{DE}$ and $P_{DE}$ are the density and pressure of dark
energy, respectively given by
\begin{eqnarray}\label{18}
\rho_{DE}&=&\frac{1}{2\kappa^2}[6H^2-F+12H^2F_T+T_GF_{T_{G}}-24H^3\dot{F}_{T_G}],\\\nonumber
P_{DE}&=&\frac{1}{2\kappa^2}[-2(2\dot{H}+3H^2)+F-4(\dot{H}+3H^2)F_T-4H\dot{F}_T-T_GF_{T_G}\\\label{19}
&+&\frac{2}{3H}T_G\dot{F}_{T_G}+8H^2\ddot{F}_{T_G}].
\end{eqnarray}
The above gravitational field equations can also be rewritten as
follows
\begin{equation}\label{20}
3H^2=\kappa^2\rho_{eff},\quad
2\dot{H}=-\kappa^2(\rho_{eff}+P_{eff}),
\end{equation}
where $\rho_{eff}=\rho_m+\rho_{DE}$ and $P_{eff}=P_m+P_{DE}$.
Furthermore, the derivatives are defined as
\begin{eqnarray}\label{21}
\dot{F}_T&=&F_TT\dot{T}+F_{TT_G}\dot{T}_G,\quad
\dot{F}_{T_G}=F_{TT_G}\dot{T}+F_{T_GT_G}\dot{T}_G, \\\nonumber
\ddot{F}_{T_G}&=&F_{TTT_G}\dot{T}^2+2F_{TT_GT_G}\dot{T}\dot{T}_G+F_{T_GT_GT_G}\ddot{T}_G^2
+F_{TT_G}\ddot{T}+F_{T_GT_G}\ddot{T}_G,\\\label{22}
\end{eqnarray}
with $F_{TT},~F_{TT_G},...$ represent the second and higher-order
differentiation with respect to $T$ and $T_G$. The energy bounds for
any modified theories of gravity are defined in terms of effective
density and pressure as follows:
\begin{eqnarray}\nonumber
\textbf{NEC}:\quad&&\rho_{eff}+P_{eff}\geq0,\\\nonumber
\textbf{SEC}:\quad&&\rho_{eff}+P_{eff}\geq0,\quad\rho_{eff}+3P_{eff}\geq0,\\\nonumber
\textbf{WEC}:\quad&&\rho_{eff}\geq0,\quad\rho_{eff}+P_{eff}\geq0,\\\label{23}
\textbf{DEC}:\quad&&\rho_{eff}\geq0,\quad\rho_{eff}\pm P_{eff}\geq0.
\end{eqnarray}
By inserting the corresponding values, the distinct inequalities in
these energy constraints in terms of $T,~T_G$ and $F$ can be written
as
\begin{eqnarray}\label{24}
&&\rho_m+\frac{1}{2\kappa^2}[6H^2-F+12H^2F_T+T_GF_{T_G}-24H^3\dot{F}_{T_G}]\geq0,\\\nonumber
&&\rho_{m}+P_{m}+\frac{1}{2\kappa^2}[-24H^3\dot{F}_{T_G}-4\dot{H}-4\dot{H}F_T-4H\dot{F}_T
+\frac{2}{3H}T_G\dot{F}_{T_G}\\\label{25}
&&+8H^2\ddot{F}_{T_G}]\geq0,
\end{eqnarray}
\begin{eqnarray}\nonumber
&&\rho_{m}+P_{m}+\frac{1}{2\kappa^2}[-12H^2+2F-2T_GF_{T_G}-24H^2F_T-24H^3\dot{F}_{T_G}-12\dot{H}F_T\\\label{26}
&&-12H\dot{F}_T+\frac{2}{H}T_G\dot{F}_{T_G}+24H^2\ddot{F}_{T_G}]\geq0,\\\nonumber
&&\rho_{m}-P_m+\frac{1}{2\kappa^2}[12H^2-2F+24H^2F_T+2T_GF_{T_G}
-24H^3\dot{F}_{T_G}+4\dot{H}+4\dot{H}F_T\\\label{27}
&&-\frac{2}{3H}T_G\dot{F}_{T_G}-8H^2\ddot{F}_{T_G}]\geq0.
\end{eqnarray}
We describe these conditions in terms of some cosmic parameters like
deceleration parameter, jerk and snap parameters defined in terms of
Hubble parameter by the following relations
\begin{equation}\nonumber
q=-(1+\frac{\dot{H}}{H^2}),\quad r=2q^2+q-\frac{\dot{q}}{H},\quad
s=\frac{(r-1)}{3(q-1/2)}.
\end{equation}
The first and higher order time rates of Hubble parameter can be
expressed in terms of these parameters by the following relations
\begin{equation}\label{28}
\dot{H}=-H^2(1+q),\quad \ddot{H}=H^3(j+3q+2),\quad
\dddot{H}=H^4(s-4j-3q(q+4)-6).
\end{equation}
The first and second-order time rates of torsion scalar $T$ and its
Gauss-Bonnet equivalent term $T_G$ are given by
\begin{eqnarray}\label{28*}
\dot{T}&=&12H\dot{H}, \quad
\dot{T}_G=24H^2(\ddot{H}+2H\dot{H})+48H\dot{H}(\dot{H}+H^2),\\\nonumber
\ddot{T}&=&12(\dot{H}^2+H\ddot{H}), \quad
\ddot{T}_G=48\dot{H}^3+144H\dot{H}\ddot{H}+288\dot{H}^2H^2+24H^2\dddot{H}\\\label{28**}
&+&96H^3\ddot{H}.
\end{eqnarray}
In terms of cosmic parameters and recent value of Hubble parameter
$H_0$, the terms $T,~T_G$ and their derivatives can be expressed as
follows
\begin{eqnarray}\label{*}
T&=&6H_0^2, \quad T_G=-24qH_0^4, \quad
\dot{T}=12H_0^3(1+q),\\\label{**}
\dot{T}_G&=&48H_0^5(1+q)^2-96H_0^5(1+q)+24H_0^5(j+3q+2),\\\label{***}
\ddot{T}&=&12H_0^4(1+q)^2+12H_0^4(j+3q+2),\\\nonumber
\ddot{T}_G&=&-48H_0^6(1+q)^3-144H_0^6(1+q)(j+3q+2)+288H_0^6(1+q)^2\\\label{****}
&+&24H_0^6(s-4j-3q(q+4)-6)+96H_0^6(j+3q+2).
\end{eqnarray}

In the following, we consider two particular models of $F(T, T_G)$
functions and discuss the corresponding constraints on the free
parameters present in these models.

\subsection{Model-1}

Here, we consider the model
\begin{equation}\label{29}
F(T,T_G)=-T+\beta_1\sqrt{T^2+\beta_2T_G}+\alpha_1T^2+\alpha_2T\sqrt{|T_G|},
\end{equation}
where $\alpha_1,~\alpha_2,~\beta_1,~\beta_2$ are free parameters to
be constrained. For this model, the derivatives can be written as
\begin{eqnarray}\nonumber
\dot{F}_T&=&[\{\beta_1(T^2+\beta_2T_G)^{-1/2}+2\alpha_1\}\dot{T}-\{\beta_1T^2\dot{T}
+\frac{1}{2}\beta_1\beta_2T\dot{T}_G\}(T^2+\beta_2T_G)^{-3/2}\\\label{30}
&+&\frac{\alpha_2}{2}(|T_G|)^{-1/2}\dot{T}_G],\\\nonumber
\dot{F}_{T_G}&=&[\{-\frac{1}{2}\beta_1\beta_2T(T^2+\beta_2T_G)^{-3/2}
+\frac{\alpha_2}{2}(|T_G|)^{-1/2}\}\dot{T}+\{-\frac{\beta_1\beta_2^2}{4}(T^2+\beta_2T_G)^{-3/2}\\\label{31}
&-&\frac{\alpha_2T}{4}(|T_G|)^{-3/2}\}\dot{T}_G],\\\nonumber
\ddot{F}_{T_G}&=&\{-\frac{\beta_1\beta_2}{2}(T^2+\beta_2T_G)^{-3/2}
+\frac{3\beta_1\beta_2}{2}T^2(T^2+\beta_2T_G)^{-5/2}\}\dot{T}^2+2\{\frac{3}{4}\beta_1\beta_2^2T(T^2\\\nonumber
&+&\beta_2T_G)^{-5/2}-\frac{\alpha_2}{4}(|T_G|)^{-3/2}\}\dot{T}\dot{T}_G
+\{\frac{3}{8}\beta_1\beta_2^3(T^2+\beta_2T_G)^{-5/2}
\\\label{32}&+&\frac{3\alpha_2T}{8}(|T_G|)^{-5/2}\}\ddot{T}_G.
\end{eqnarray}
Inserting all these values in energy constraints
(\ref{24})-(\ref{27}), we get the following inequalities:
\begin{eqnarray}\nonumber
&&\rho_{eff}+P_{eff}\geq0 \Rightarrow
\rho^m+P^m+\frac{1}{2\kappa^2}[(\frac{2T_G}{3H}-24H^3)\{(-\frac{1}{2}\beta_1\beta_2T
(T^2+\beta_2T_G)^{-3/2}\\\nonumber
&&+\frac{\alpha_2}{2}(|T_G|)^{-1/2})\dot{T}+(-\frac{\beta_1\beta_2^2}{4}
(T^2+\beta_2T_G)^{-3/2}-\frac{\alpha_2T}{4}(|T_G|)^{-3/2})\dot{T}_G\}\\\nonumber
&&-4\dot{H}-4\dot{H}\{-1+\beta_1T(T^2+\beta_2T_G)^{-1/2}+2\alpha_1T+\alpha_2(|T_G|)^{1/2}\}\\\nonumber
&&-4H\{(\beta_1(T^2+\beta_2T_G)^{-1/2})-\beta_1T^2(T^2+\beta_2T_G)^{-3/2}+2\alpha_1)\dot{T}\\\nonumber
&&+(-\frac{1}{2}\beta_1\beta_2T(T^2+\beta_2T_G)^{-3/2}+\frac{\alpha_2}{2}(|T_G|)^{-1/2})\dot{T}_G\}\\\nonumber
&&+8H^2\{(-\frac{1}{2}\beta_1\beta_2(T^2+\beta_2T_G)^{-3/2}+\frac{3}{2}\beta_1\beta_2T^2(T^2+\beta_2T_G)^{-5/2})\dot{T}^2\\\nonumber
&&+2(\frac{3}{4}\beta_1\beta_2^2T(T^2+\beta_2T_G)^{-3/2}-\frac{\alpha_2}{4}(|T_G|)^{-3/2})\dot{T}\dot{T}_G\\\label{33}
&&+(\frac{3}{8}\beta_1\beta_2^3(T^2+\beta_2T_G)^{-3/2}+\frac{3\alpha_2T}{8}(|T_G|)^{-5/2})\ddot{T}_G\}]\geq0,\\\nonumber
&&\rho_{eff}\geq0\Rightarrow\rho^m+\frac{1}{2\kappa^2}[6H^2-(-T+\beta_1(T^2+\beta_2T_G)^{1/2}\\\nonumber
&&+\alpha_1T^2+\alpha_2T(|T_G|)^{1/2})+12H^2\{-1+\beta_1T(T^2+\beta_2T_G)^{-1/2}+2\alpha_1T\\\nonumber
&&+\alpha_2(|T_G|)^{1/2}\}+T_G\{\frac{1}{2}\beta_1\beta_2(T^2+\beta_2T_G)^{-1/2}+\frac{\alpha_2T}{2}(|T_G|)^{-1/2}\}\\\nonumber
&&-24H^3\{(-\frac{1}{2}\beta_1\beta_2T(T^2+\beta_2T_G)^{-3/2}+\frac{\alpha_2}{2}(|T_G|)^{-1/2})\dot{T}\\\label{34}
&&+(-\frac{\beta_1\beta_2^2}{4}(T^2+\beta_2T_G)^{-3/2}-\frac{\alpha_2T}{4}(|T_G|)^{-3/2})\dot{T}_G\}]\geq0,\\\nonumber
&&\rho_{eff}+3P_{eff}\geq0\Rightarrow\rho^m+3P^m+\frac{1}{2\kappa^2}[-12H^2+2\{-T+\beta_1(T^2+\beta_2T_G)^{1/2}\\\nonumber
&&+\alpha_1T^2+\alpha_2T(|T_G|)^{1/2}\}-2T_G\{\frac{\beta_1\beta_2}{2}(T^2+\beta_2T_G)^{-1/2}\\\nonumber
&&+\frac{\alpha_2T}{2}(|T_G|)^{-1/2}\}-(24H^2+12\dot{H})\{-1+\beta_1T(T^2+\beta_2T_G)^{-1/2}\\\nonumber
&&+2\alpha_1T+\alpha_2(|T_G|)^{1/2}\}-12H\{(\beta_1(T^2+\beta_2T_G)^{-1/2}-\beta_1T^2(T^2+\beta_2T_G)^{-3/2}+2\alpha_1)\dot{T}\\\nonumber
&&+(-\frac{1}{2}\beta_1\beta_2T(T^2+\beta_2T_G)^{-3/2}+\frac{\alpha_2}{2}(|T_G|)^{-1/2})\dot{T}_G\}\\\nonumber
&&+(\frac{2}{H}T_G-24H^3)\{(-\frac{1}{2}\beta_1\beta_2T(T^2+\beta_2T_G)^{-3/2}+\frac{\alpha_2}{2}(|T_G|)^{-1/2})\dot{T}
\end{eqnarray}
\begin{eqnarray}\nonumber\\\nonumber
&&+(-\frac{\beta_1\beta_2^2}{4}(T^2+\beta_2T_G)^{-3/2}-\frac{\alpha_2T}{4}(|T_G|)^{-3/2})\dot{T}_G\}\\\nonumber
&&+24H^2\{(-\frac{\beta_1\beta_2}{2}(T^2+\beta_2T_G)^{-3/2}+\frac{3}{2}\beta_1\beta_2T^2(T^2+\beta_2T_G)^{-5/2})\dot{T}^2\\\nonumber
&&+2(\frac{3}{4}\beta_1\beta_2^2T(T^2+\beta_2T_G)^{-5/2}-\frac{\alpha_2}{4}(|T_G|)^{-3/2})\dot{T}\dot{T}_G\\\label{35}
&&+(\frac{3}{8}\beta_1\beta_2^3(T^2+\beta_2T_G)^{-5/2}+\frac{3}{8}\alpha_2T(|T_G|)^{-5/2})\ddot{T}_G\}]\geq0,\\\nonumber
&&\rho_{eff}-P_{eff}\geq0\Rightarrow\rho^m-P^m+\frac{1}{2\kappa^2}[12H^2-2\{-T+\beta_1(T^2+\beta_2T_G)^{1/2}\\\nonumber
&&+\alpha_1T^2+\alpha_2T(|T_G|)^{1/2}\}+(24H^2+4\dot{H})\{-1+\beta_1T(T^2+\beta_2T_G)^{-1/2}+2\alpha_1T\\\nonumber
&&+\alpha_2(|T_G|)^{1/2}\}+2T_G\{\frac{\beta_1\beta_2}{2}(T^2+\beta_2T_G)^{-1/2}+\frac{\alpha_2T}{2}(|T_G|)^{-1/2}\}\\\nonumber
&&-(24H^3+\frac{2T_G}{3H})\{(-\frac{1}{2}\beta_1\beta_2T(T^2+\beta_2T_G)^{-3/2}+\frac{\alpha_2}{2}(|T_G|)^{-1/2})\dot{T}\\\nonumber
&&+(-\frac{\beta_1\beta_2^2}{4}(T^2+\beta_2T_G)^{-3/2}-\frac{\alpha_2T}{4}(|T_G|)^{-3/2})\dot{T}_G\}\\\nonumber
&&+4\dot{H}+4H\{(\beta_1(T^2+\beta_2T_G)^{-1/2}-\beta_1T^2(T^2+\beta_2T_G)^{-3/2}+2\alpha_1)\dot{T}\\\nonumber
&&+(-\frac{1}{2}\beta_1\beta_2(T^2+\beta_2T_G)^{-3/2}+\frac{\alpha_2}{2}(|T_G|)^{-1/2})\dot{T}_G\}\\\nonumber
&&-8H^2\{(-\frac{1}{2}\beta_1\beta_2(T^2+\beta_2T_G)^{-3/2}+\frac{3}{2}\beta_1\beta_2T^2(T^2+\beta_2T_G)^{-3/2})\dot{T}^2\\\nonumber
&&+2(\frac{3}{4}\beta_1\beta_2^2T(T^2+\beta_2T_G)^{-5/2}-\frac{\alpha_2}{4}(|T_G|)^{-3/2})\dot{T}\dot{T}_G\\\label{36}
&&+(\frac{3}{8}\beta_1\beta_2^3(T^2+\beta_2T_G)^{-5/2}+\frac{3\alpha_2T}{8}(|T_G|)^{-5/2})\ddot{T}_G\}]\geq0.
\end{eqnarray}
Here, each of these inequalities depend on the values of
$\rho^m,~P^m,~\beta_1,~\beta_2,\alpha_1,~\alpha_2$
$T_G,~T,~\alpha_1,~\alpha_2,\dot{T},~\dot{T}_G,~\dot{H},~H$ and
$\ddot{T}_G$. Firstly, we discuss these constraints using recent
values of cosmic parameters like $H_0$, $q_0,~j_0$ and $s_0$. For
this purpose, we use the values proposed by Capozziello et al.
\cite{36}. These values are $H_0=0.718,~q_0=-0.64,~j_0=1.02$ and
$s_0=-0.39$. We assume that the energy constraints are satisfied for
ordinary matter quantities. Also, $\kappa^2=\frac{8\pi G}{c^4}$ is
gravitational coupling constant and hence a positive quantity,
therefore we only investigate the inequalities for DE source and
find the possible constraints on the free parameters
$\alpha_1,~\alpha_2,~\beta_1$ and $\beta_2$ by fixing any two of
them. In terms of present day values, the above inequalities
(\ref{33})-(\ref{36}) result in the following set of constraints
given by
\begin{eqnarray}\nonumber
&&\rho_{eff}=\{(-7.510\beta_1(-7.215+\beta_2)(1.689+\beta_2)+\sqrt{9.568
+4.082\beta_2}(\alpha_2\\\nonumber&&\times(59.004+25.175\beta_2)+(274.613+117.168\beta_2)))/\{9.568
+4.082\beta_2\}^{\frac{3}{2}}\geq0,\\\nonumber &&\rho_{eff}+p_{eff}=
(-4.633\beta_1(-144.801+\beta_2)(0.137+\beta_2)(2.279+\beta_2)+
(229.584\alpha_1\\\nonumber&&+53.403\alpha_2)(2.344+\beta_2)^2
\sqrt{9.568+4.0822\beta_2}\}/\{9.568+4.0822\beta_2\}^{\frac{5}{2}}\geq0,\\\nonumber
&&\rho_{eff}+3p_{eff}=\{9.568+4.082\beta_2)^{\frac{5}{2}}\}\{60.834\beta_1
(-0.239+\beta_2)(2.285+\beta_2)\\\nonumber&&\times(33.696+\beta_2)+\sqrt{9.568+4.082\beta_2}
(-37.112(2.344+\beta_2)^2-267.847\alpha_1\\\nonumber&&\times(2.344+\beta_2)^2-259.101\alpha_2
(2.344+\beta_2)(2.344+\beta_2))\}\geq0,\\\nonumber
&&\rho_{eff}-p_{eff}=\{-70.101\beta_1(-0.889+\beta_2)(2.293+\beta_2)(10.790+\beta_2)
\\\nonumber&&+(1.480\times10^{-14}+727.015\alpha_1+365.907\alpha_2)(2.344+\beta_2)^2
\\\nonumber&&\times\sqrt{9.568+4.082\beta_2}\}/\{9.568+4.082\beta_2\}^{\frac{5}{2}}\geq0.
\end{eqnarray}
We present the numerical evolution of above inequalities in terms of
constants $\alpha_1$, $\alpha_2$, $\beta_1$ and $\beta_2$. In right
Figure \textbf{1}, we show the variation of $\rho_{eff}\geq0$ versus
parameters $\alpha_1$ and $\alpha_2$. In right panel, we set
$\beta_1=0.1$ and $\beta_2=0.2$ and it is found that WEC can be
satisfied for all positive values of $\alpha_1$ and $\alpha_2$. In
left panel, we set the negative values of $\beta_i(i=1,2)$, which
requires $\beta_2\leq-2$. In this case, the constraint of WEC can be
met if $\alpha_i(i=1,2)>0$. In Figure \textbf{2}, we fix $\alpha_i$
to determine the possible constraints on the parameters $\beta_1$
and $\beta_2$. In left panel, we explore the evolution for
$\alpha=0.1$ and $\alpha_2=0.2$, it can be seen that WEC is
satisfied only if $0\leqslant\beta_i\leqslant7$. We also explore the
variation of WEC for negative values of $\alpha_i$ in right plot and
find similar results.
\begin{figure}
\center\epsfig{file=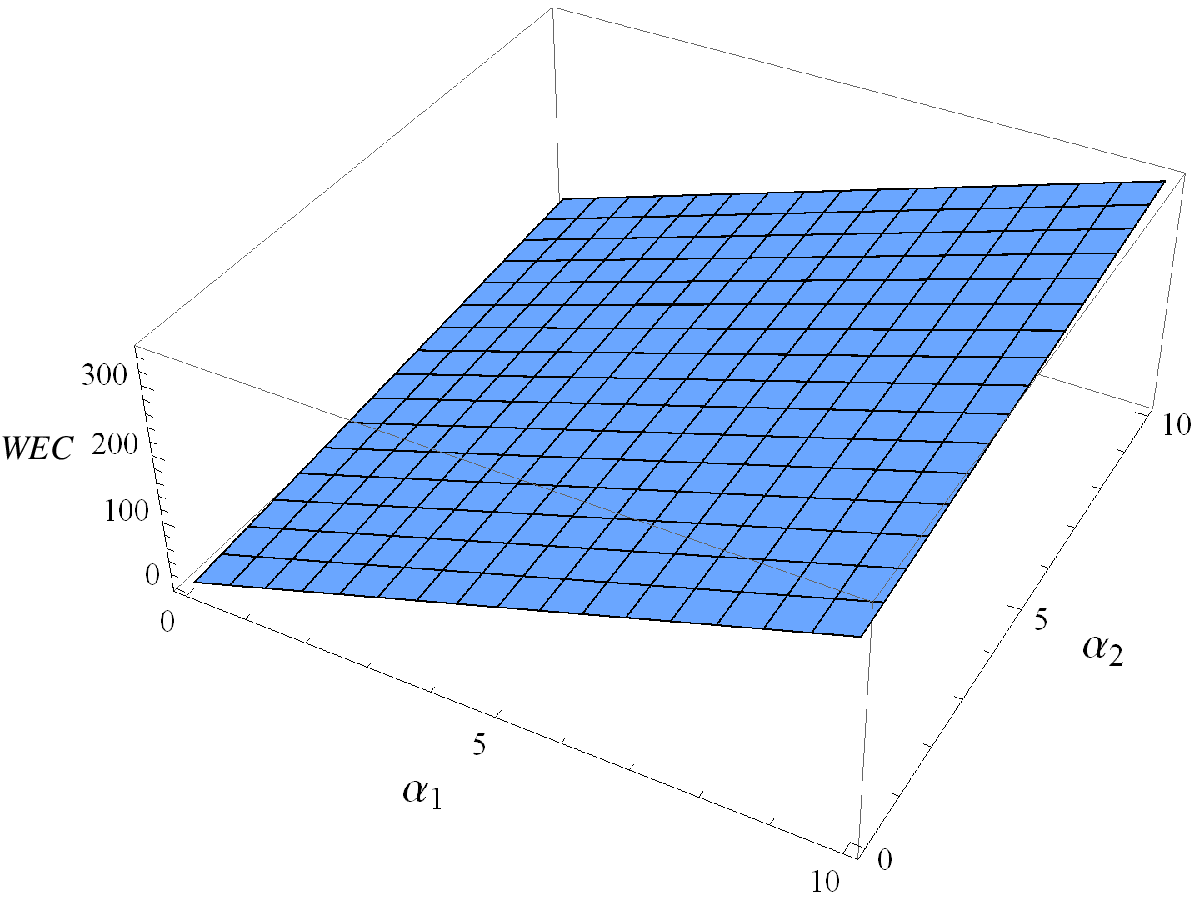, width=0.45\linewidth}
\epsfig{file=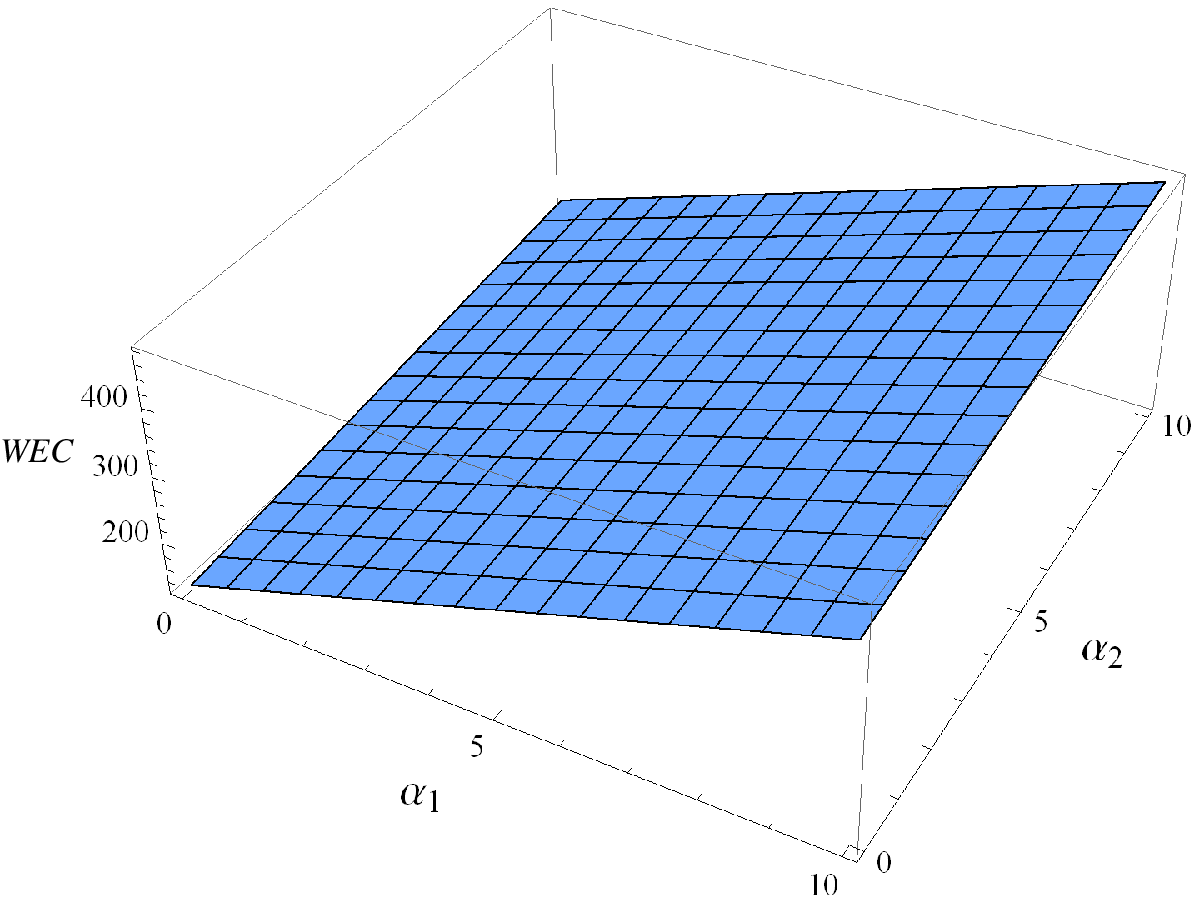, width=0.45\linewidth}\caption{Evolution of
WEC versus the parameters $\alpha_i (i=1,2)$ and $\beta_i (i=1,2)$.
The left plot corresponds to parameters $\beta_1=0.1$ and
$\beta_2=0.2$ and right plot corresponds to $\beta_1=-10$ and
$\beta_2=-2$.}
\end{figure}
\begin{figure}
\center\epsfig{file=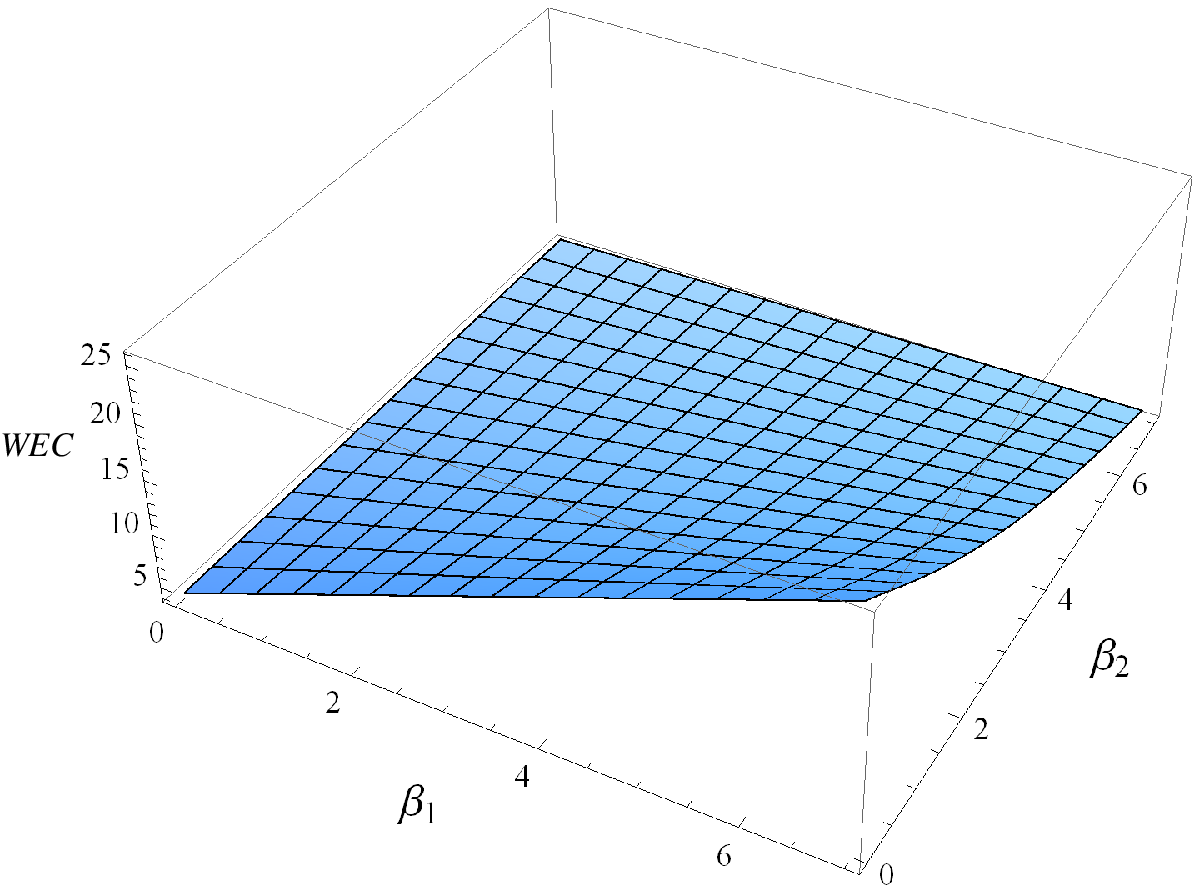, width=0.45\linewidth}
\epsfig{file=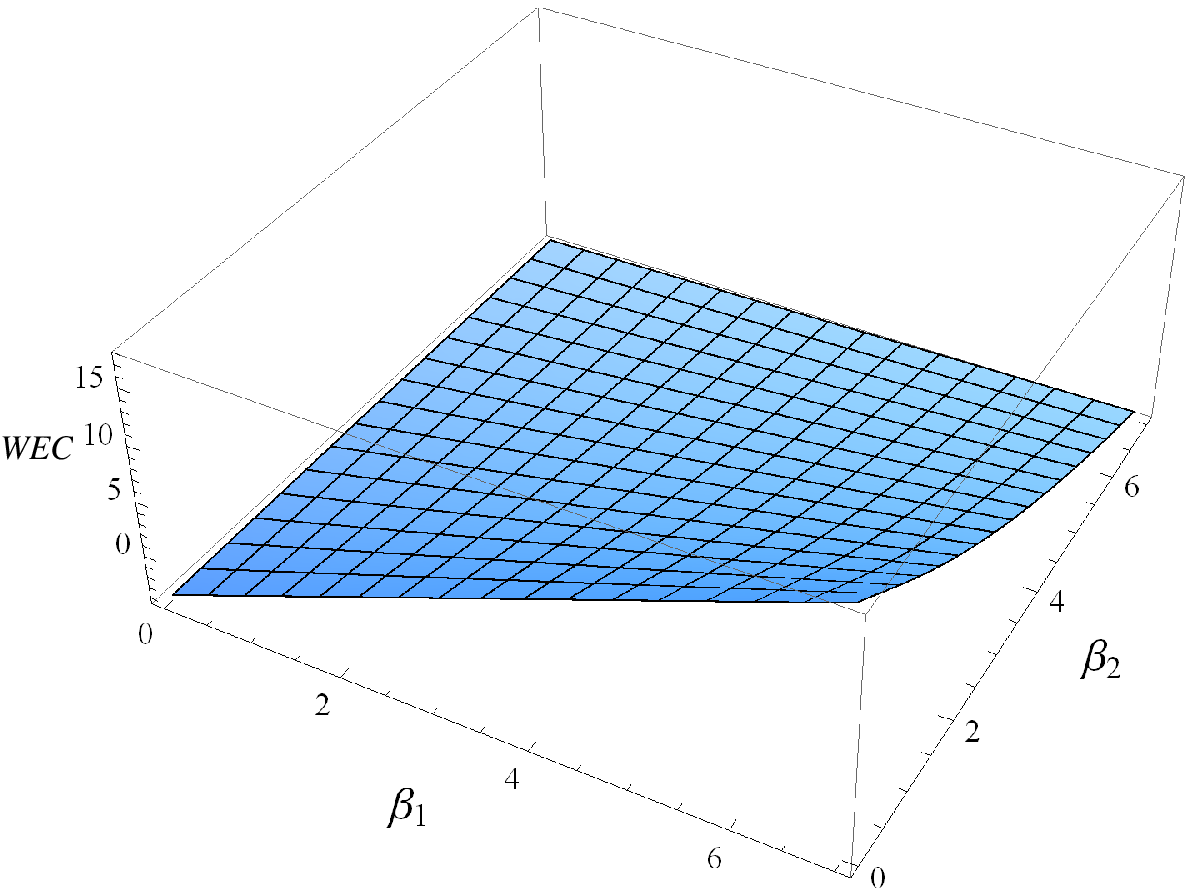, width=0.45\linewidth} \caption{Evolution of
WEC versus the parameters $\alpha_i (i=1,2)$ and $\beta_i (i=1,2)$.
The left plot corresponds to parameters $\alpha_1=0.1$ and
$\alpha_2=0.2$ and right plot corresponds to $\alpha_1=-0.1$ and
$\alpha_2=-0.2$.}
\end{figure}

The plots of NEC versus the parameters $\alpha_i$ and $\beta_i$ for
model (\ref{29}) are shown in Figure \textbf{3}. In left plot, we
fix $\beta_1=0.1$ and $\beta_2=0.2$ to show the variation for
$\alpha_i$, whereas in right plot, we show the variation of
$\beta_i$ for $\alpha_1=0.1$ and $\alpha_2=0.2$. It is interesting
to mention here that one can get similar results for negative values
of $\alpha_i$ and $\beta_i$. In case of SEC, we present the
variations for free parameters in Figure \textbf{4} and discussed
the difference in results. In left plot, we set $\beta_1=0.1$ and
$\beta_2=0.2$, to determine the evolution of SEC versus $\alpha_i$.
It can be seen that SEC is satisfied if $\alpha_i<0$. We also
explore the variation of SEC versus $\beta_i$ which is found to be
independent of signature of $\alpha_i$. For the discussion of
possible constraints arising due to DEC, we explore its evolution in
Figure \textbf{5}. In left plot, we show that DEC is satisfied only
if $\alpha_i>0$ for all values of $\beta_i$. For $\alpha_1=0.1$ and
$\alpha_2=0.2$, DEC can be met if $\beta_1>0$ and
$\beta_2\geqslant-2$ as shown in right plot.
\begin{figure}
\center\epsfig{file=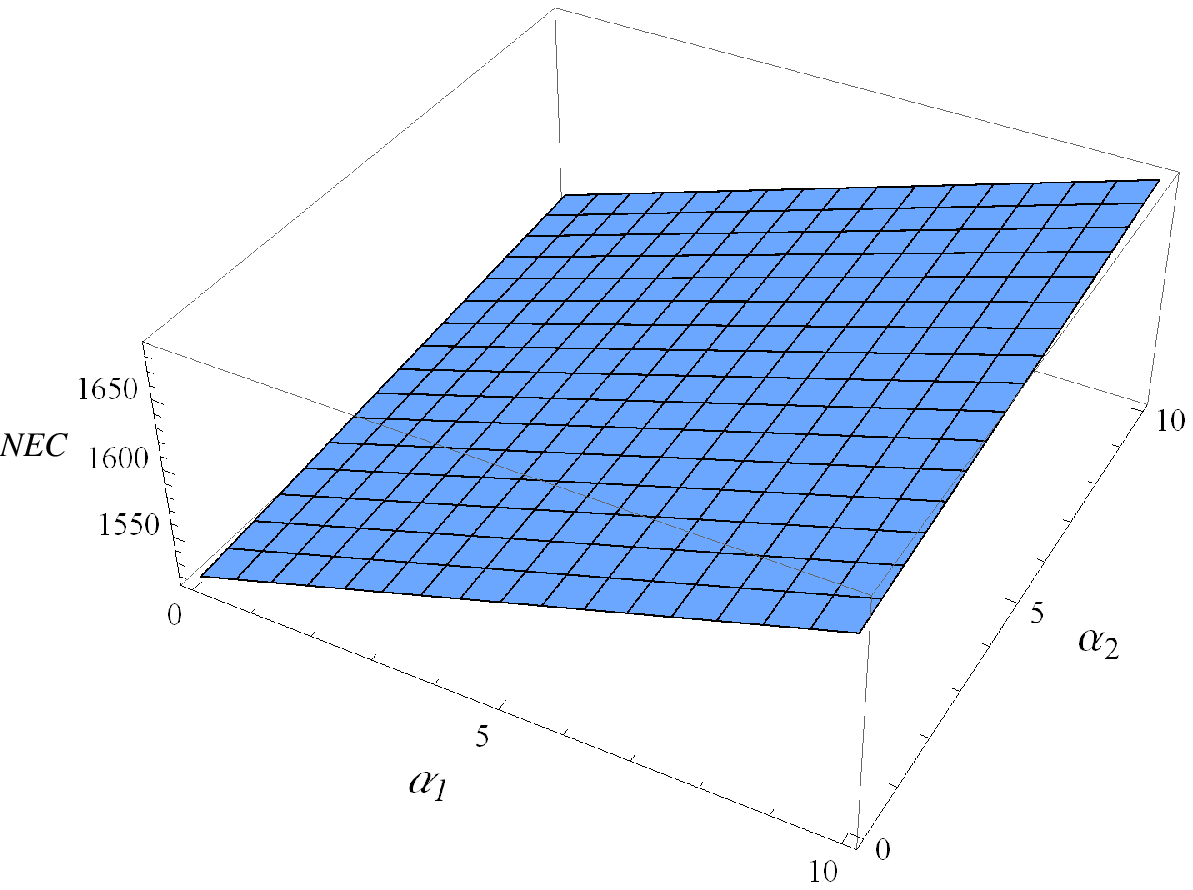, width=0.45\linewidth}
\epsfig{file=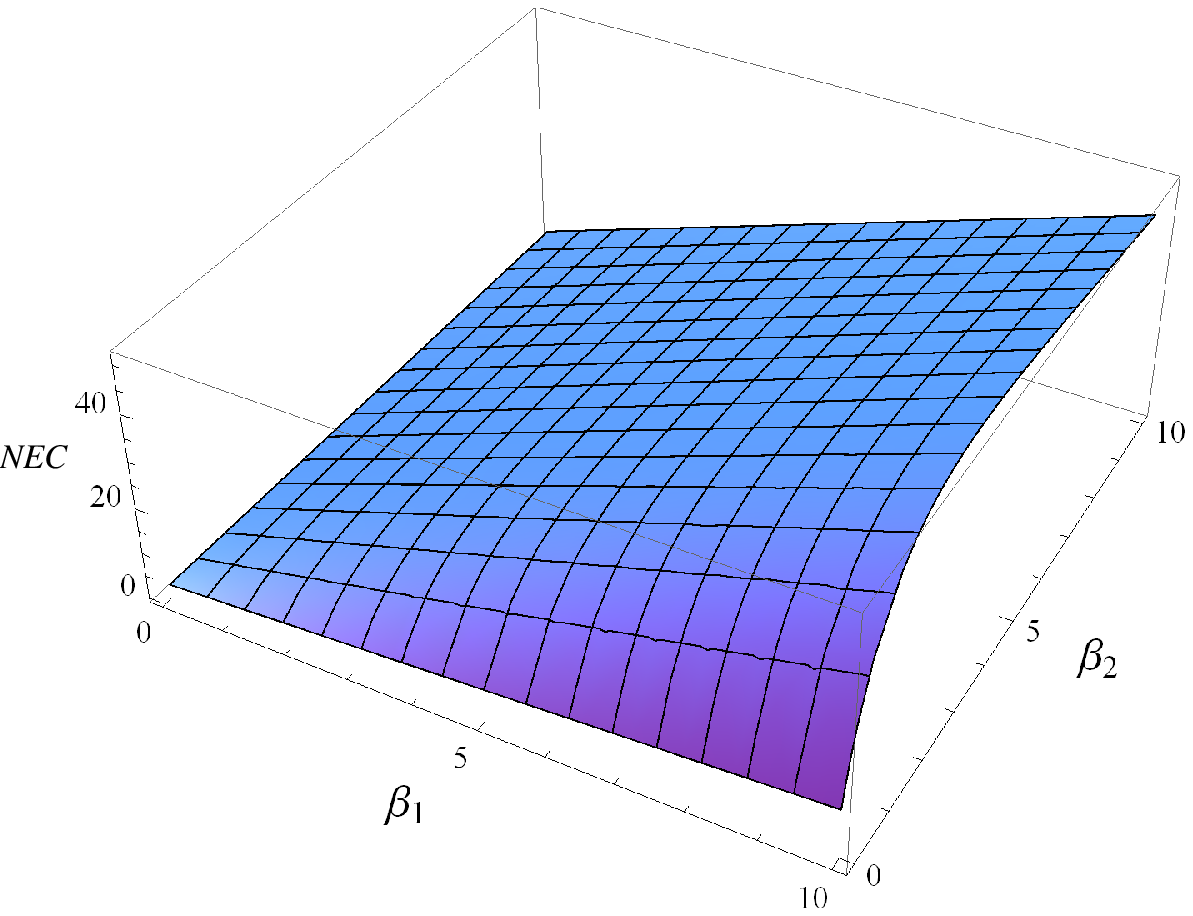, width=0.45\linewidth} \caption{Evolution of
NEC versus the parameters $\alpha_i (i=1,2)$ and $\beta_i (i=1,2)$.
The left plot corresponds to $\beta_1=0.1$ and $\beta_2=0.2$ and
right plot corresponds to $\alpha_1=0.1$ and $\alpha_2=0.2$.}
\end{figure}
\begin{figure}
\center\epsfig{file=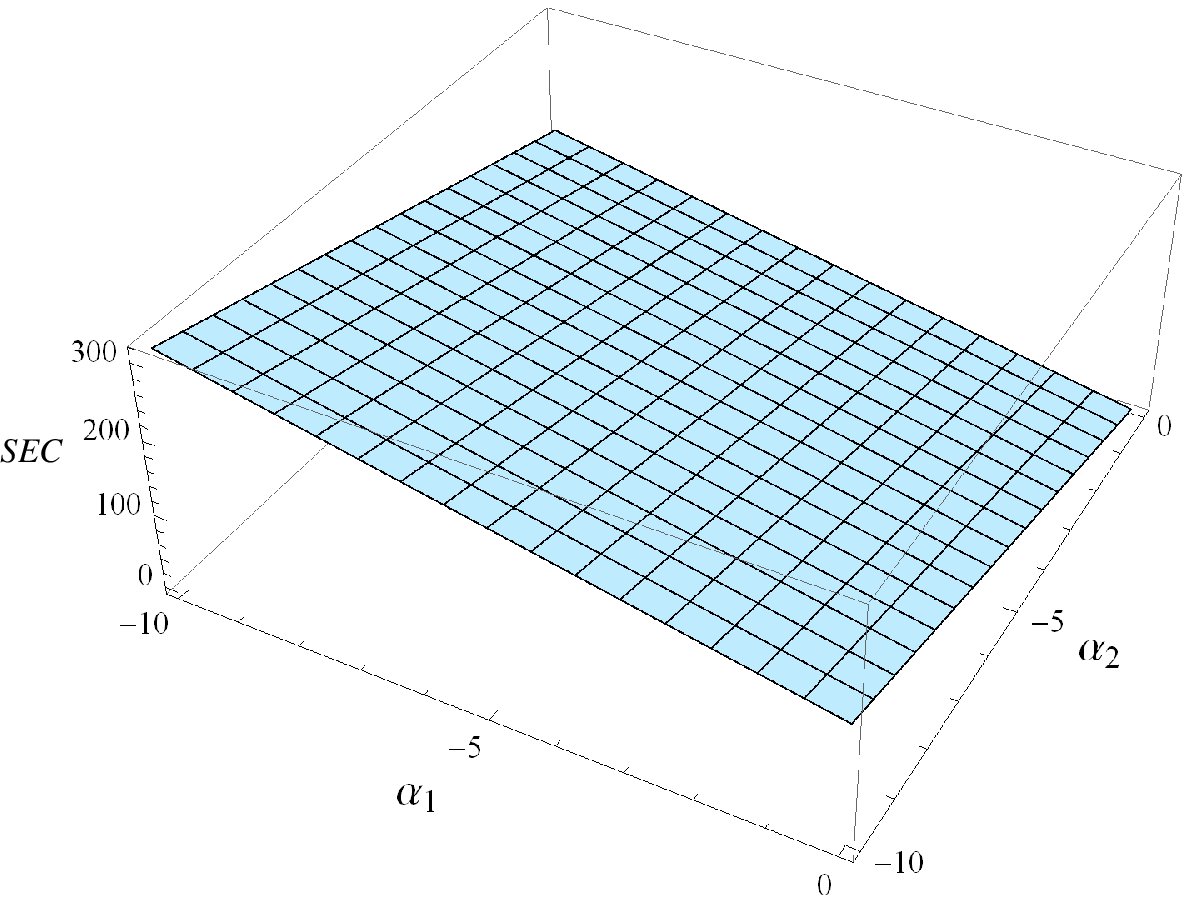, width=0.45\linewidth}
\epsfig{file=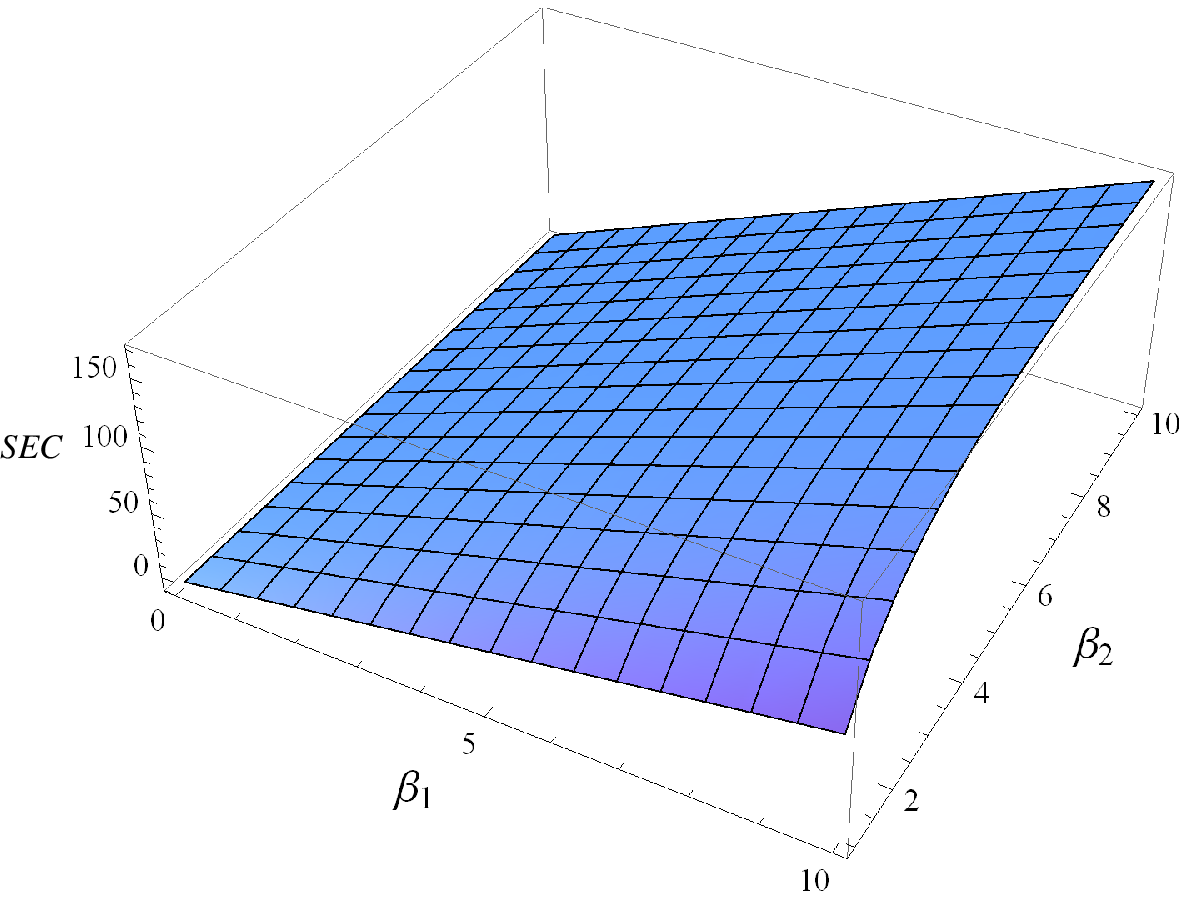, width=0.45\linewidth} \caption{Evolution of
SEC versus the parameters $\alpha_i (i=1,2)$ and $\beta_i (i=1,2)$.
The left plot corresponds to $\beta_1=0.1$ and $\beta_2=0.2$ and
right plot corresponds to $\alpha_1=0.1$ and $\alpha_2=0.2$.}
\end{figure}
\begin{figure}
\center\epsfig{file=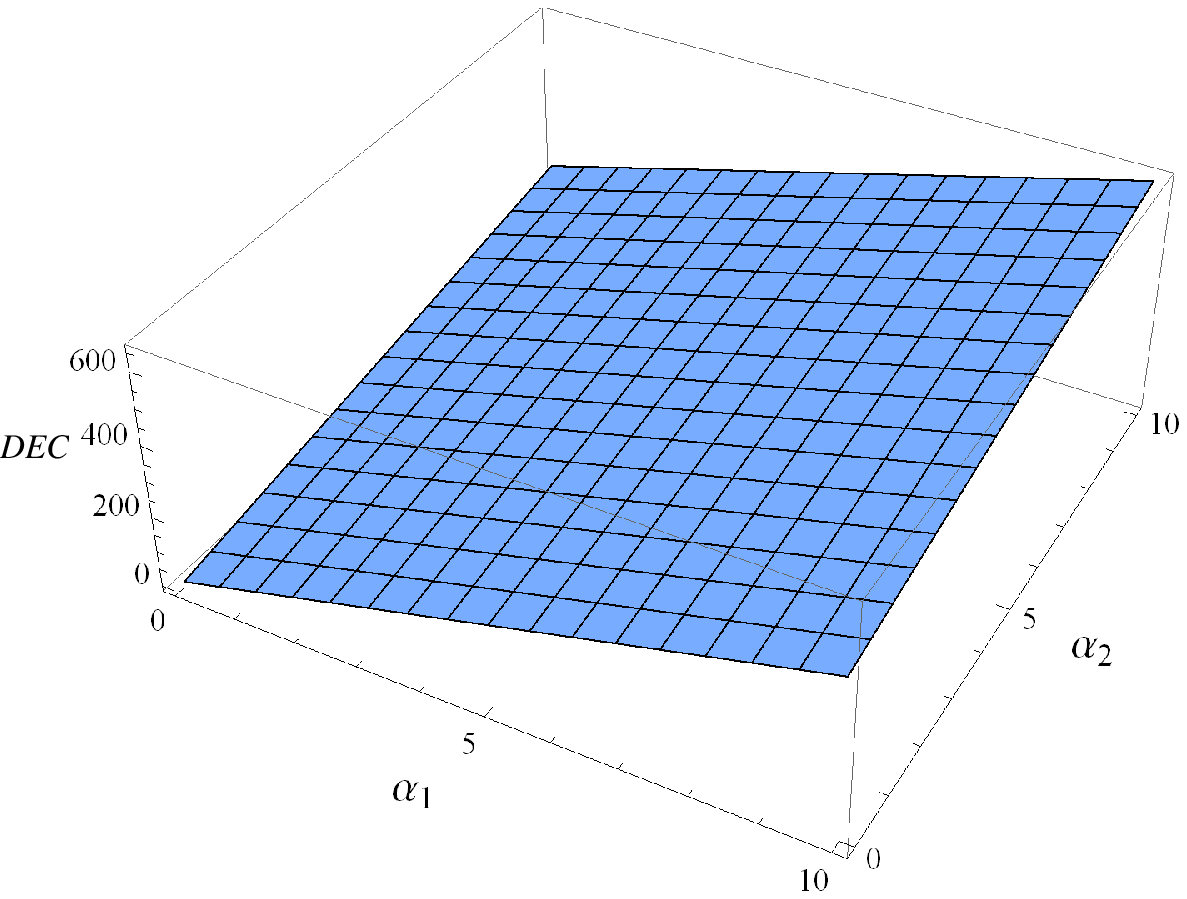, width=0.45\linewidth}
\epsfig{file=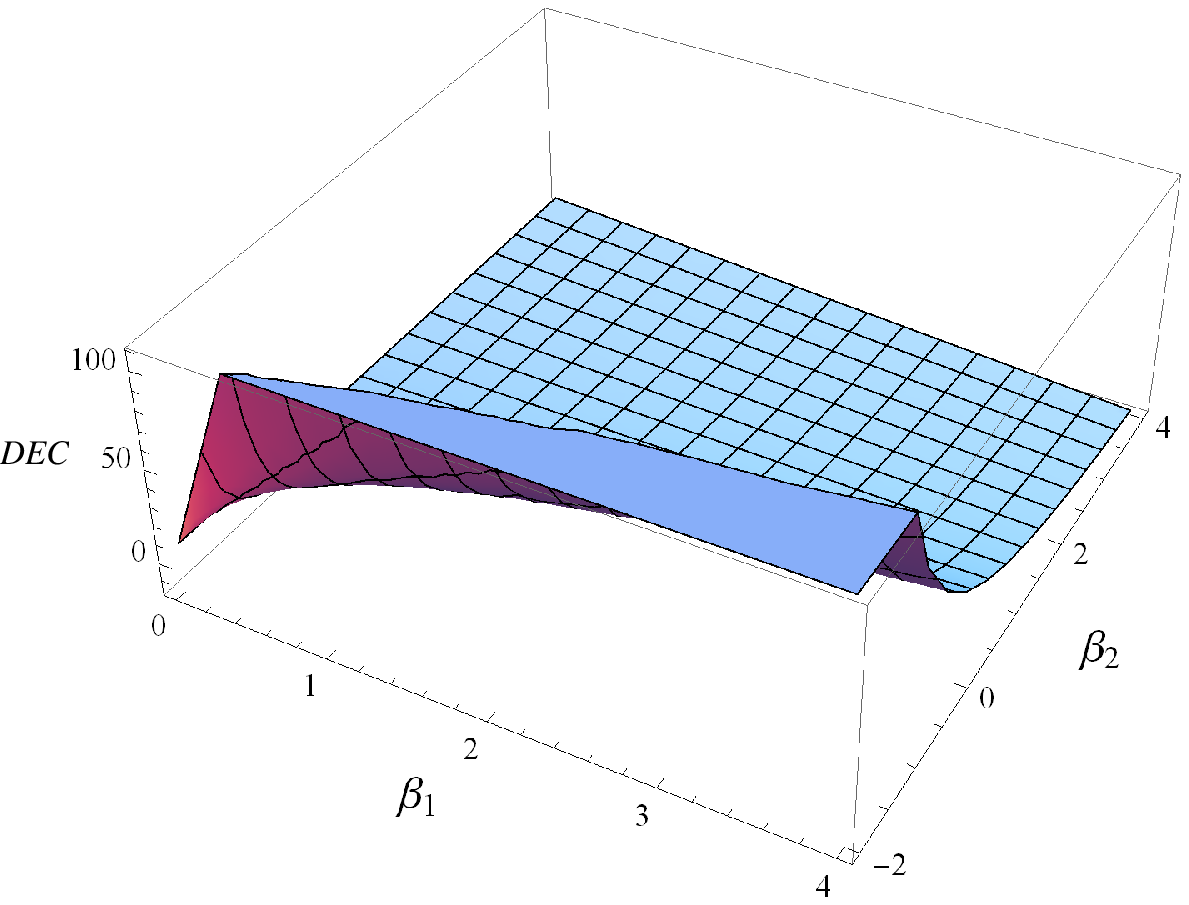, width=0.45\linewidth} \caption{Evolution of
DEC versus the parameters $\alpha_i (i=1,2)$ and $\beta_i (i=1,2)$.
The left plot corresponds to $\beta_1=0.1$ and $\beta_2=0.2$ and
right plot corresponds to $\alpha_1=0.1$ and $\alpha_2=0.2$.}
\end{figure}

The power law cosmology is described by $a(t)=a_0t^m$, where $m$ is
any arbitrary constant which further leads to $H=\frac{m}{t}$. It is
interesting to mention here that for $0<m<1$, the power law
cosmology corresponds to decelerating universe, while for the values
satisfying $m>1$, this leads to accelerating cosmic model. The
first, second and third-order time rates of Hubble parameter are
$$\dot{H}=-\frac{m}{t^2}, \quad \ddot{H}=\frac{2m}{t^3}, \quad \dddot{H}=-\frac{6m}{t^4}.$$
We can also discuss the time rates of $T$ and $T_G$ in terms of
cosmic time as follows
\begin{eqnarray}\label{*****}
T&=&\frac{6m^2}{t^2}, \quad T_G=24\frac{m^3(m-1)}{t^4}, \quad
\dot{T}=-\frac{12m^2}{t^3},\\\label{******}
\dot{T}_G&=&-\frac{96m^3(m-1)}{t^5}, \quad
\ddot{T}=\frac{36m^2}{t^4}, \quad
\ddot{T}_G=-\frac{480m^3}{t^6}-\frac{144m^4}{t^6}.
\end{eqnarray}
One can find the energy constraints in power law cosmology for this
theory by using the above defined relations. In Figures \textbf{6}
and \textbf{7}, we explore the evolution of constraints arising from
weak, null, strong and dominant energy conditions in terms of $m$
and $t$ with $\alpha_1=10$, $\alpha_2=0.2$, $\beta_1=2$,
$\beta_2=1$. In case of SEC, one need to set $\alpha_1=0.001$. We
find that energy constraints can be satisfied for all values of $m$
and $t$.
\begin{figure}
\center\epsfig{file=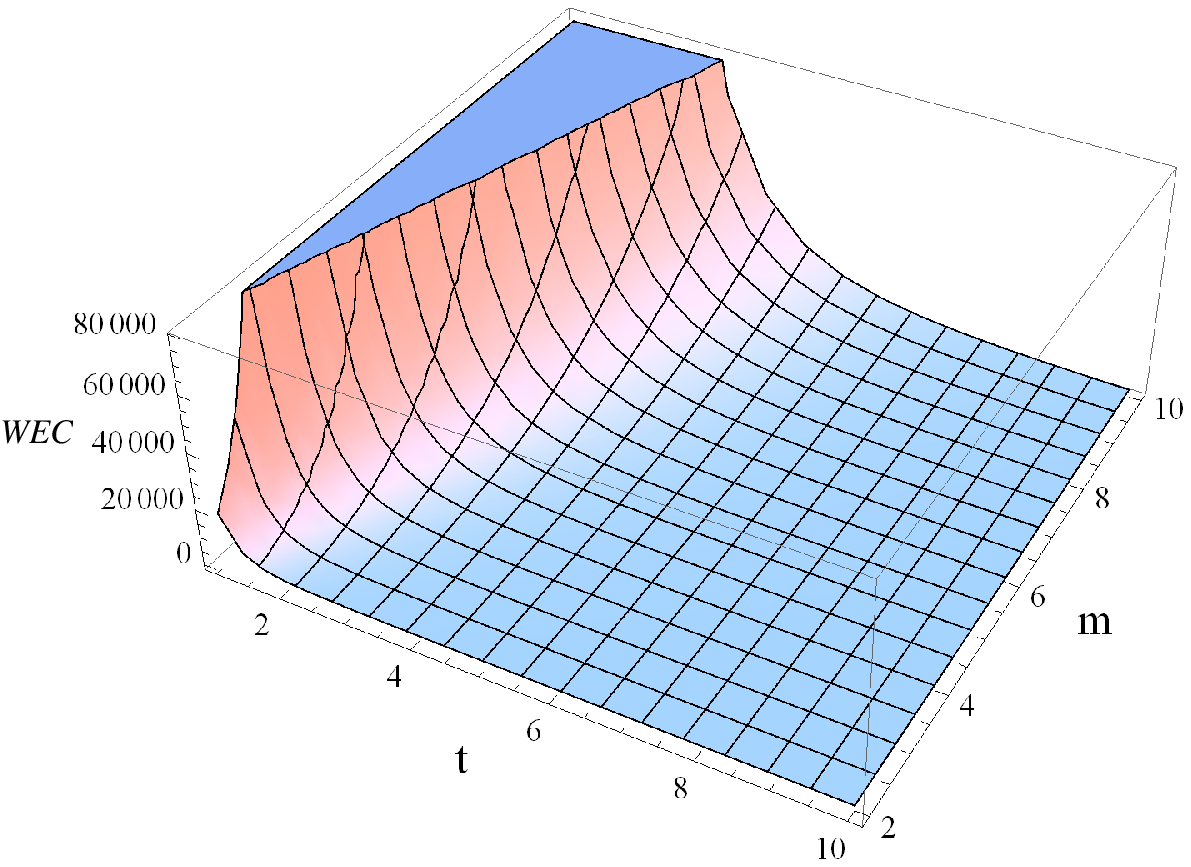, width=0.45\linewidth}
\epsfig{file=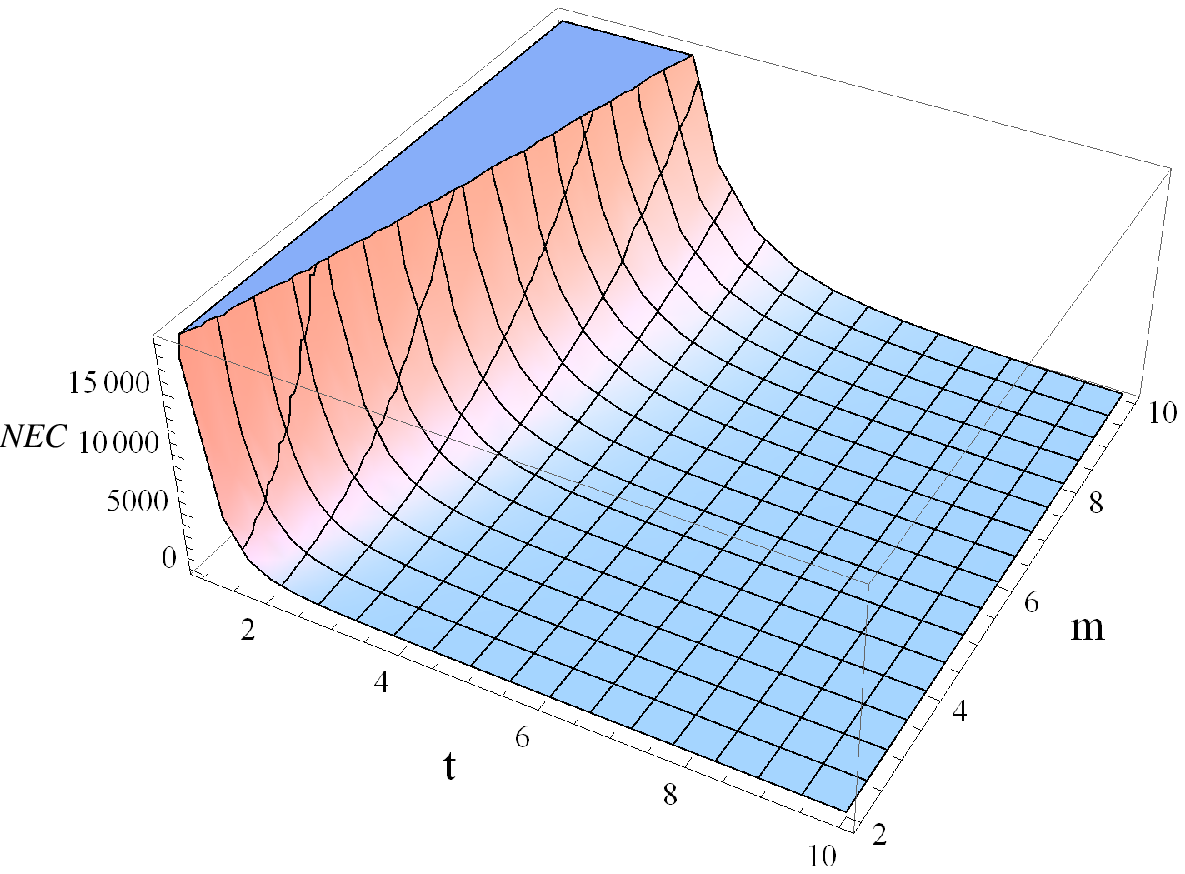, width=0.45\linewidth} \caption{Evolution of
WEC and NEC versus $m$ and $t$ with $\alpha_1=10$, $\alpha_2=0.2$,
$\beta_1=2$, $\beta_2=1$.}
\end{figure}
\begin{figure}
\center\epsfig{file=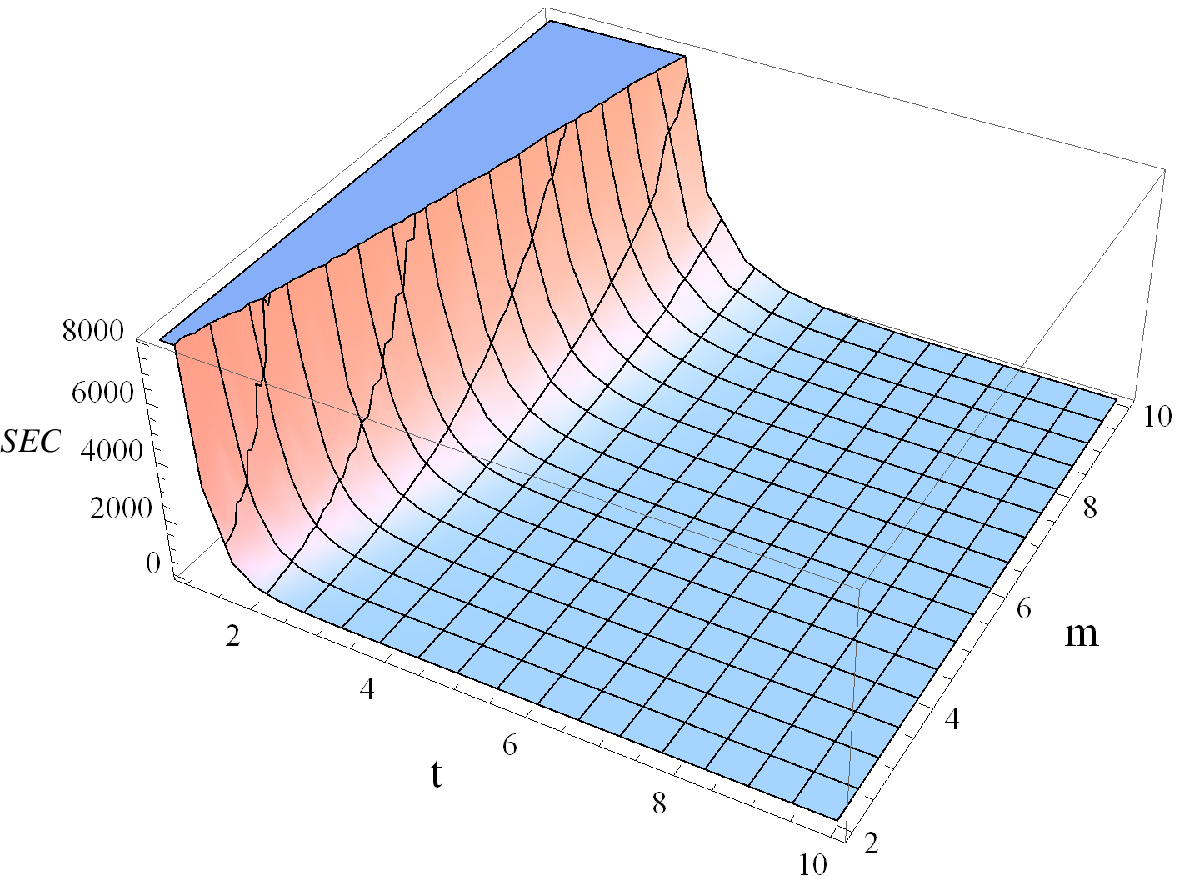, width=0.45\linewidth}
\epsfig{file=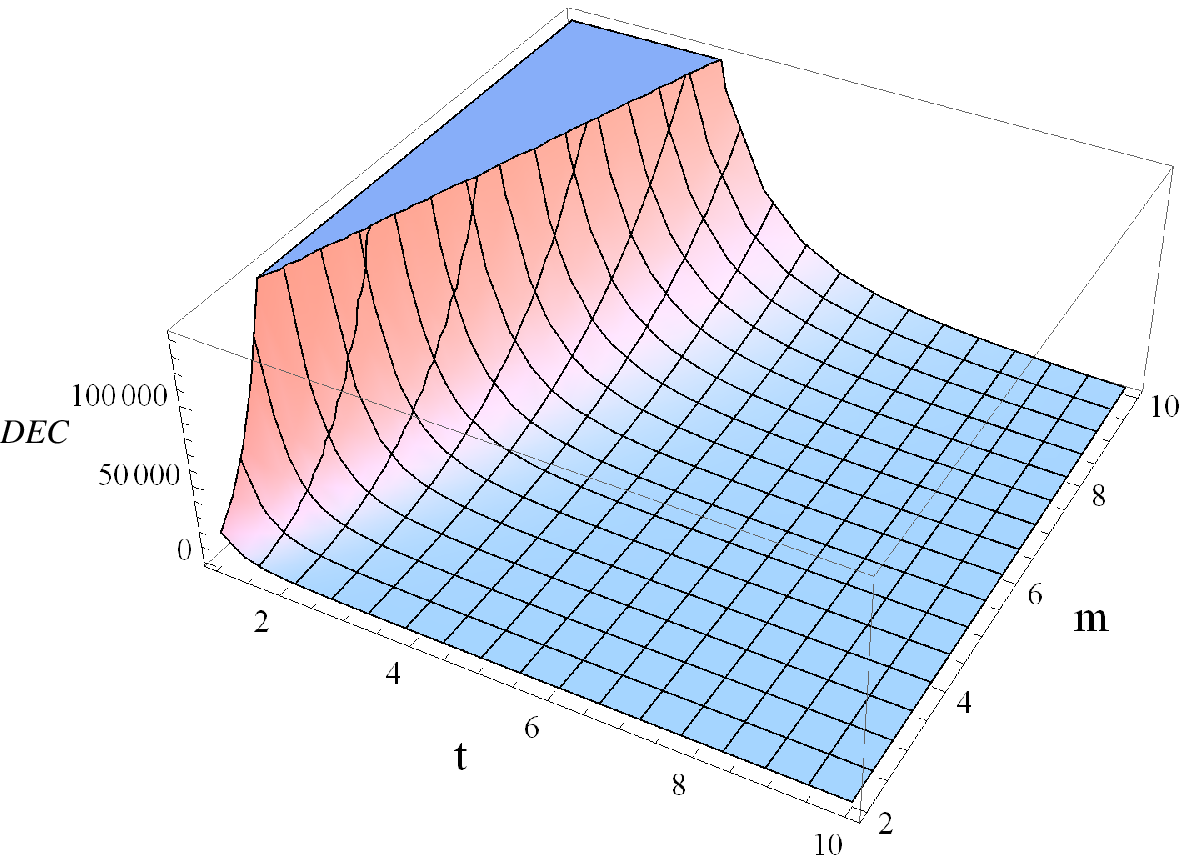, width=0.45\linewidth} \caption{Evolution of
SEC and DEC versus $m$ and $t$ with $\alpha_1=10$, $\alpha_2=0.2$,
$\beta_1=2$, $\beta_2=1$.}
\end{figure}

\subsection{Model-2}

In this section, we consider the following form of $F$ given by
\begin{equation}\label{29*}
F(T,T_G)=-T+\beta_1(T^2+\beta_2T_G)+\beta_3(T^2+\beta_4T_G)^2,
\end{equation}
where $\beta_1,~\beta_2,~\beta_3$ and $\beta_4$ are the free
parameters to be constrained. For this model, the derivatives of $F$
can be written as
\begin{eqnarray}\nonumber
\dot{F}_T&=&(2\beta_1+12\beta_3T^2+4\beta_3\beta_4T_G)\dot{T}+4\beta_3\beta_4
T\dot{T}_G,\\\nonumber
\dot{F}_{T_G}&=&4\beta_3\beta_4T\dot{T}+2\beta_3\beta_4^2\dot{T}_G,\\\nonumber
\ddot{F}_{T_G}&=&4\beta_3\beta_4\dot{T}^2+4\beta_3\beta_4T\ddot{T}+2\beta_3\beta_4^2\ddot{T}_G.
\end{eqnarray}
For this model, the energy constraints take the following form
\begin{eqnarray}\nonumber
&&\rho_{eff}+P_{eff}\geq0\Rightarrow
\rho^m+P^m+\frac{1}{2\kappa^2}[-24H^3(4\beta_3\beta_4T\dot{T}+2\beta_3\beta_4^2\dot{T}_G)\\\nonumber
&&-4\dot{H}-4\dot{H}(-1+2\beta_1T+4\beta_3T(T^2+\beta_4T_G))-4H\{(2\beta_1+12\beta_3T^2\\\nonumber
&&+4\beta_3\beta_4T_G)\dot{T}+4\beta_3\beta_4T\dot{T}_G\}+\frac{2T_G}{3H}\{4\beta_3\beta_4T\dot{T}+2\beta_3\beta_4^2\dot{T}_G\}
+8H^2\{4\beta_3\beta_4\dot{T}^2\\\label{}
&&+4\beta_3\beta_4T\ddot{T}+2\beta_3\beta_4^2\ddot{T}_G\}]\geq0,\\\nonumber
&&\rho_{eff}\geq0\Rightarrow\rho^m+\frac{1}{2\kappa^2}[6H^2-(-T+\beta_1(T^2+\beta_2T_G)+\beta_3(T^2+\beta_4T_G)^2)\\\nonumber
&&+12H^2(-1+2\beta_1T+4\beta_3T(T^2+\beta_4T_G))+T_G(\beta_1\beta_2+2\beta_3\beta_4(T^2+\beta_4T_G))\\\label{}
&&-24H^3(4\beta_3\beta_4T\dot{T}+2\beta_3\beta_4^2\dot{T}_G)]\geq0,\\\nonumber
&&\rho_{eff}+3P_{eff}\geq0\Rightarrow\rho^m+3P^m+\frac{1}{2\kappa^2}[-12H^2+2(-T+\beta_1(T^2+\beta_2T_G)\\\nonumber
&&+\beta_3(T^2+\beta_4T)^2)-2T_G(\beta_1\beta_2+2\beta_3\beta_4(T^2+\beta_4T_G))-(24H^2+12\dot{H})\\\nonumber
&&\times(-1+2\beta_1T+4\beta_3T(T^2+\beta_4T_G))-(24H^3-\frac{2T_G}{H})\{4\beta_3\beta_4T\dot{T}+2\beta_3\beta_4^2\dot{T}_G\}\\\nonumber
&&-12H\{(2\beta_1+12\beta_3T^2+4\beta_3\beta_4T_G)\dot{T}+4\beta_3\beta_4T\dot{T}_G\}+24H^2\{4\beta_3\beta_4\dot{T}^2\\\label{}
&&+4\beta_3\beta_4T\ddot{T}+2\beta_3\beta_4^2\ddot{T}_G\}]\geq0,\\\nonumber
&&\rho_{eff}-P_{eff}\geq0\Rightarrow\rho^m-P^m+\frac{1}{2\kappa^2}[12H^2-2(-T+\beta_1(T^2+\beta_2T_G)\\\nonumber
&&+\beta_3(T^2+\beta_4T_G))+(24H^2+4\dot{H})(-1+2\beta_1T+4\beta_3T(T^2+\beta_4T_G))\\\nonumber
&&+2T_G(\beta_1\beta_2+2\beta_3\beta_4(T^2+\beta_4T_G))+4\dot{H}+4H\{(2\beta_1+12\beta_3T^2+4\beta_3\beta_4T_G)\dot{T}\\\nonumber
&&+4\beta_3\beta_4^2\dot{T}_G\}-(24H^3+\frac{2T_G}{3H})\{4\beta_3\beta_4T\dot{T}+2\beta_3\beta_4^2\dot{T}_G\}
-8H^2\{4\beta_3\beta_4\dot{T}^2\\\label{}
&&+4\beta_3\beta_4T\ddot{T}+2\beta_3\beta_4^2\ddot{T}_G\}]\geq0.
\end{eqnarray}
Here each of these inequalities depend upon the values
$\rho^m,~P^m,~H,~\kappa^2,~T,~T_G,$
$\dot{H},~\dot{T},\dot{T}_G,~\ddot{T}_G$ and
$\beta_1,~\beta_2,~\beta_3,~\beta_4$ except the inequality given by
Eq.(\ref{}) which does not depend upon the free parameter $\beta_2$.
In terms of present day values of cosmic parameters $H$, $q$, $j$
and $s$, the above inequalities result in
\begin{eqnarray}\label{48}
\rho_{eff}&=&28.703\beta_1+ 640.765\beta_3+435.343\beta_3\beta_4+
23.238\beta_3\beta_4^2\geq0,\\\nonumber \rho_{eff}+p_{eff}&=&
13.777\beta_1+87.876\beta_3+393.686\beta_3\beta_4+8609.38\beta_3^2\beta_4-
10.685\\\label{49}&&\times\beta_3\beta_4^2\geq0,\\\nonumber
\rho_{eff}+3p_{eff}&=&
-2.227-16.074\beta_1-1017.9\beta_3+204.654\beta_3\beta_4+25828.1\beta_3^2
\\\label{50}&&\times\beta_4-78.530\beta_3\beta_4^2\geq0,\\\nonumber
\rho_{eff}-p_{eff}&=&
8.88178\times10^{-16}+43.628\beta_1+1193.65\beta_3+582.718\beta_3\beta_4\\\label{51}&&-
8609.38\beta_3^2\beta_4+57.1613\beta_3\beta_4^2\geq0.
\end{eqnarray}
Equations (\ref{48})-(\ref{51}) show energy constraints in terms of
$\beta_i$'s, obtained through the recent valued cosmic parameters.
We show the evolution of these inequalities in Figures \textbf{8}
and \textbf{9}. In these plots, we fix $\beta_1=0.1$ and show
variations against $\beta_3$ and $\beta_4$. Here WEC can be
satisfied if $\beta_3>0$ and $\beta_4>0$, whereas in case of other
constraints arising from NEC, SEC and DEC, one can set
$\beta_3\gtrless0$. We also explore the energy constraints in terms
of power law solution and set the parameters in a way to examine the
evolution against $m$ and $t$. Figures \textbf{10} and \textbf{11}
indicate that in power law cosmology, energy conditions for model
(\ref{29*}) can be satisfied in terms of $m$ and $t$ for particular
values of parameters $\beta_i$.
\begin{figure}
\center\epsfig{file=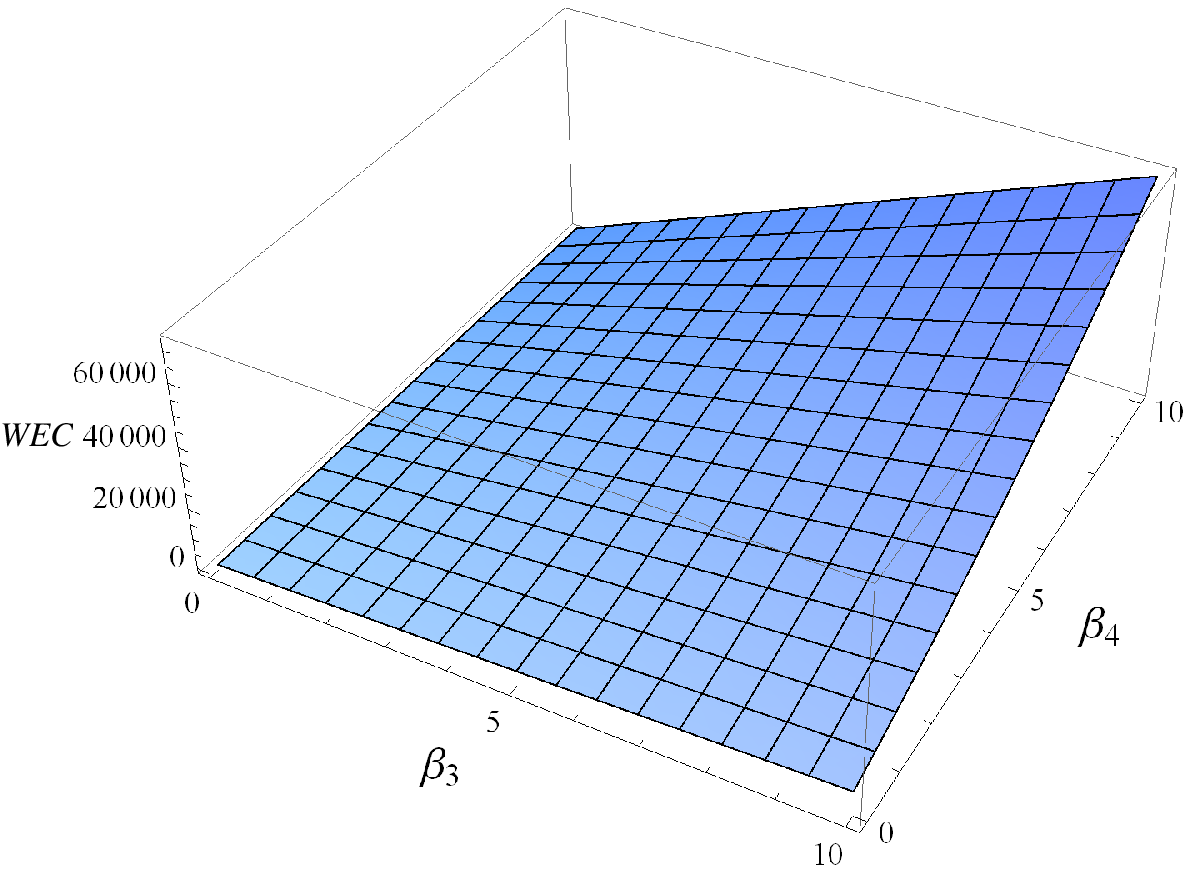, width=0.45\linewidth}
\epsfig{file=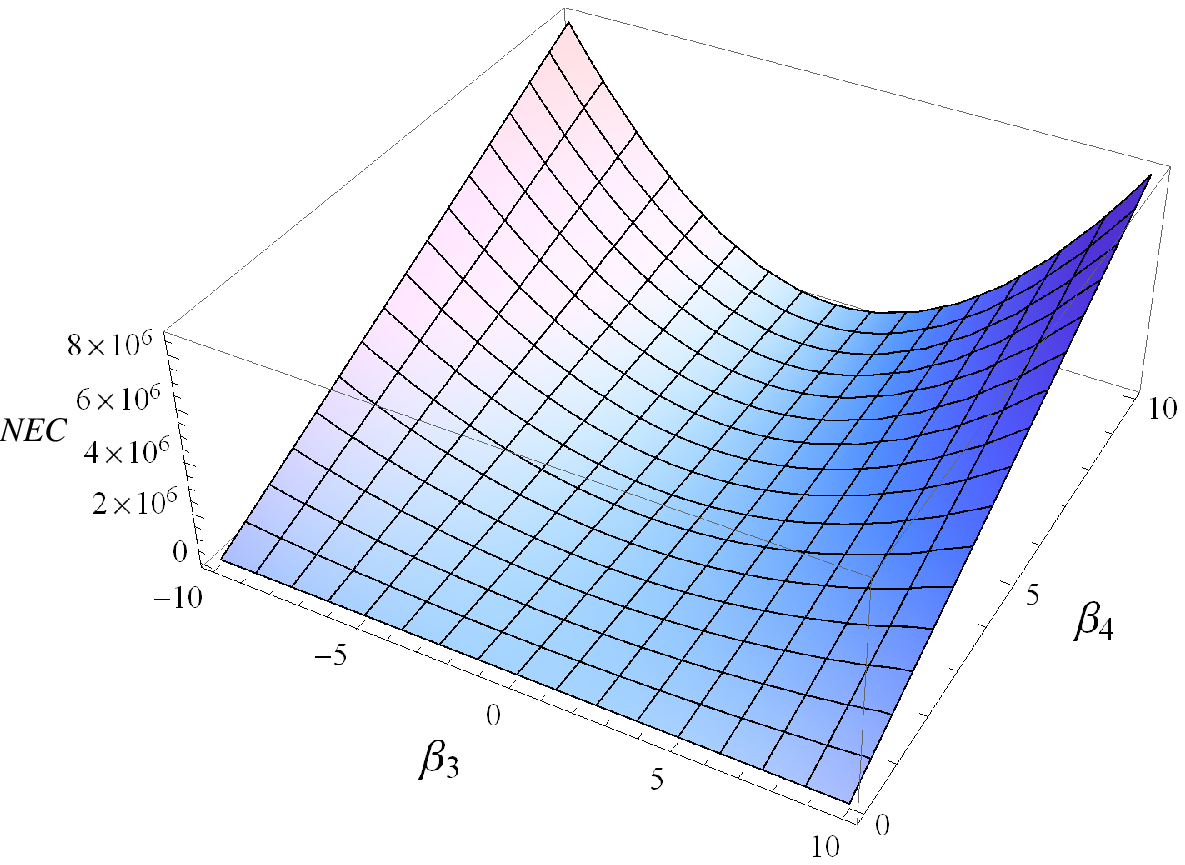, width=0.45\linewidth} \caption{Evolution of
WEC and NEC versus $\beta_3$, $\beta_4$ with $\beta_1=0.1$.}
\end{figure}
\begin{figure}
\center\epsfig{file=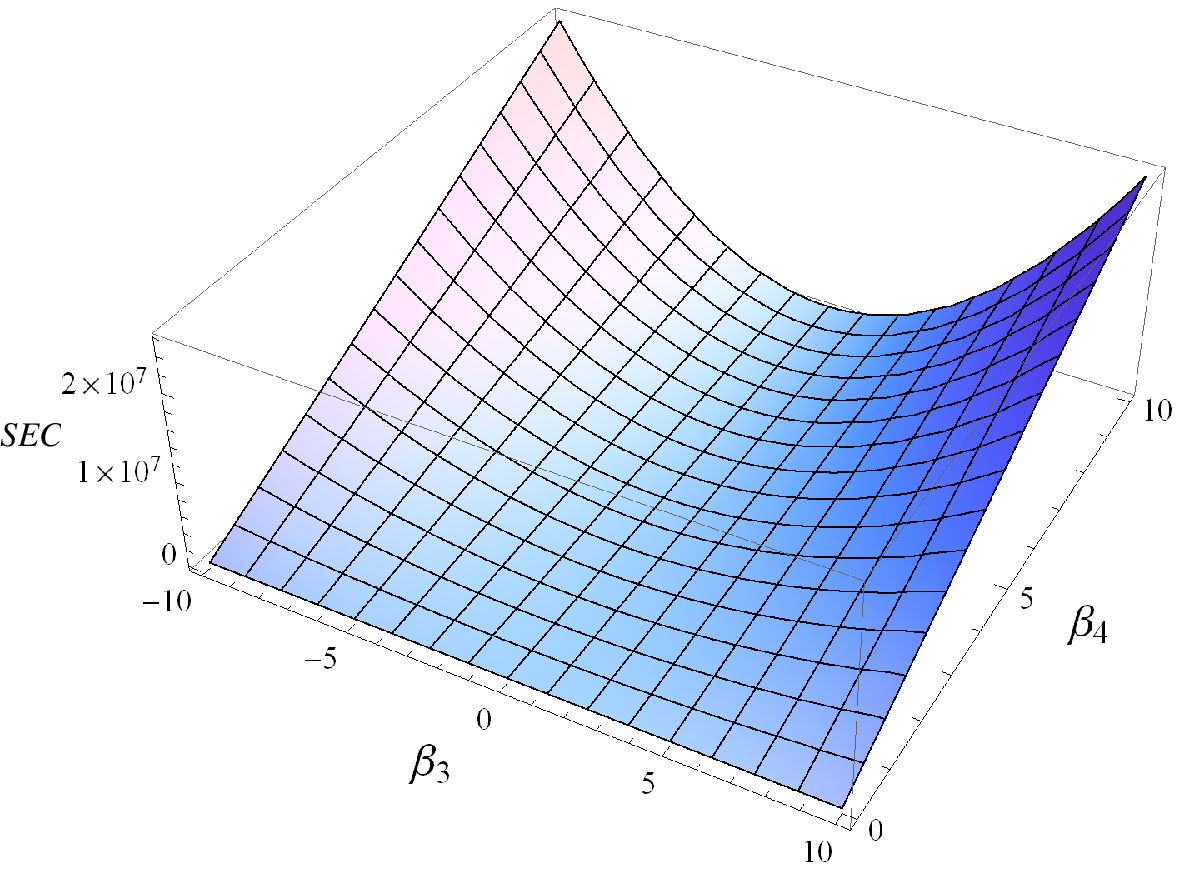, width=0.45\linewidth}
\epsfig{file=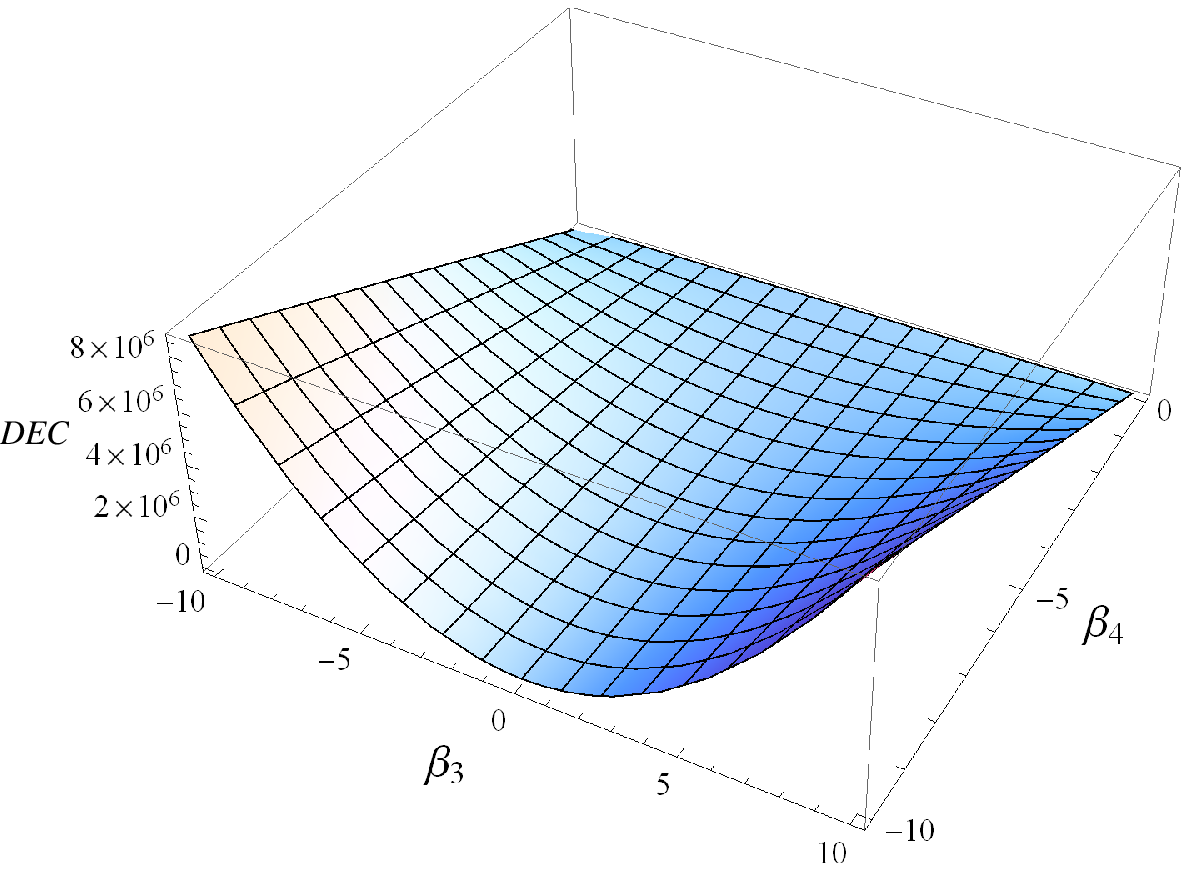, width=0.45\linewidth} \caption{Evolution of
SEC and DEC versus $\beta_3$, $\beta_4$ with $\beta_1=0.1$.}
\end{figure}
\begin{figure}
\center\epsfig{file=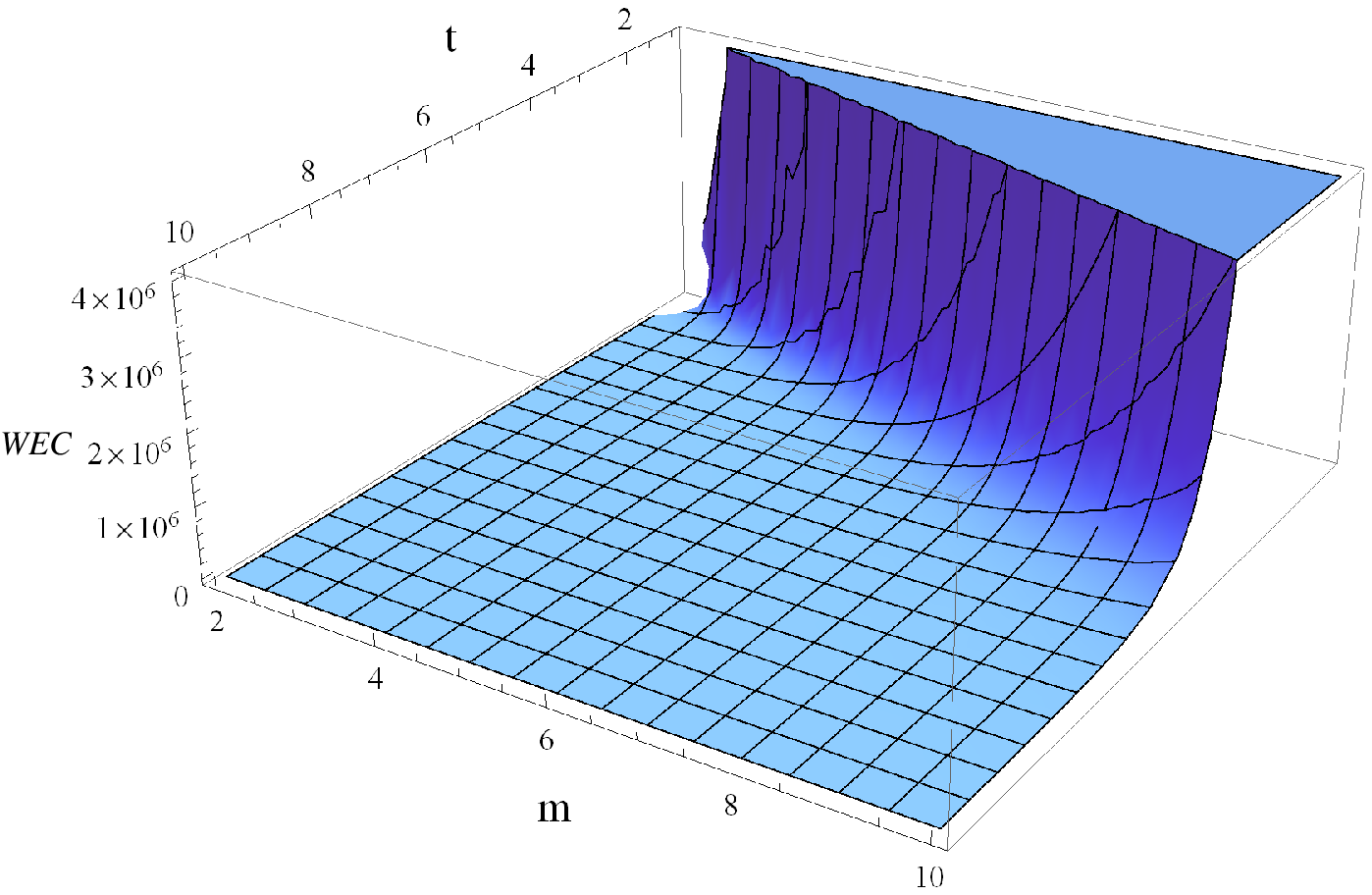, width=0.45\linewidth}
\epsfig{file=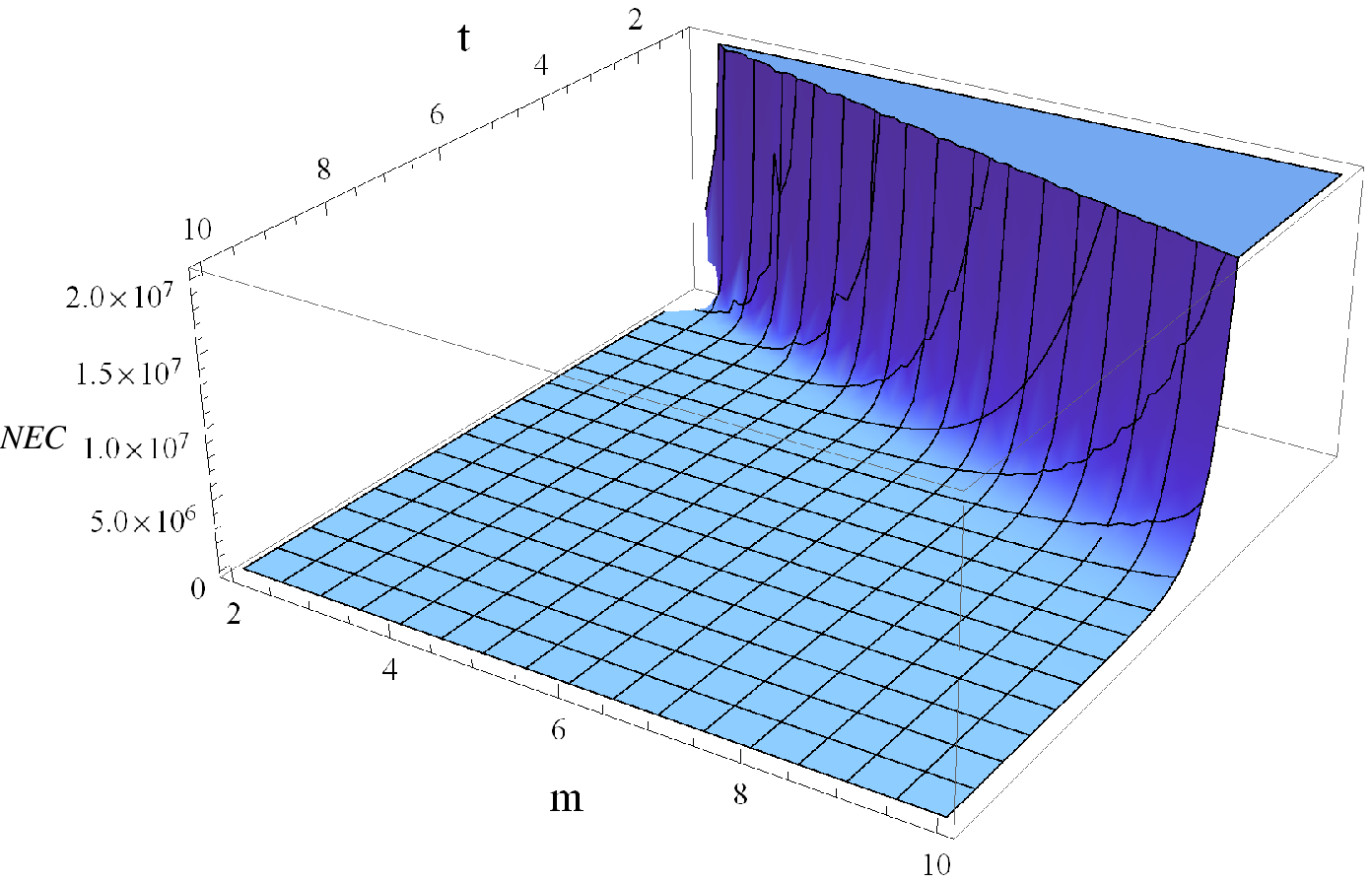, width=0.45\linewidth} \caption{Evolution
of WEC and NEC versus $m$ and $t$ with $\beta_1=0.001$,
$\beta_3=0.2$, $\beta_4=0.03$.}
\end{figure}
\begin{figure}
\center\epsfig{file=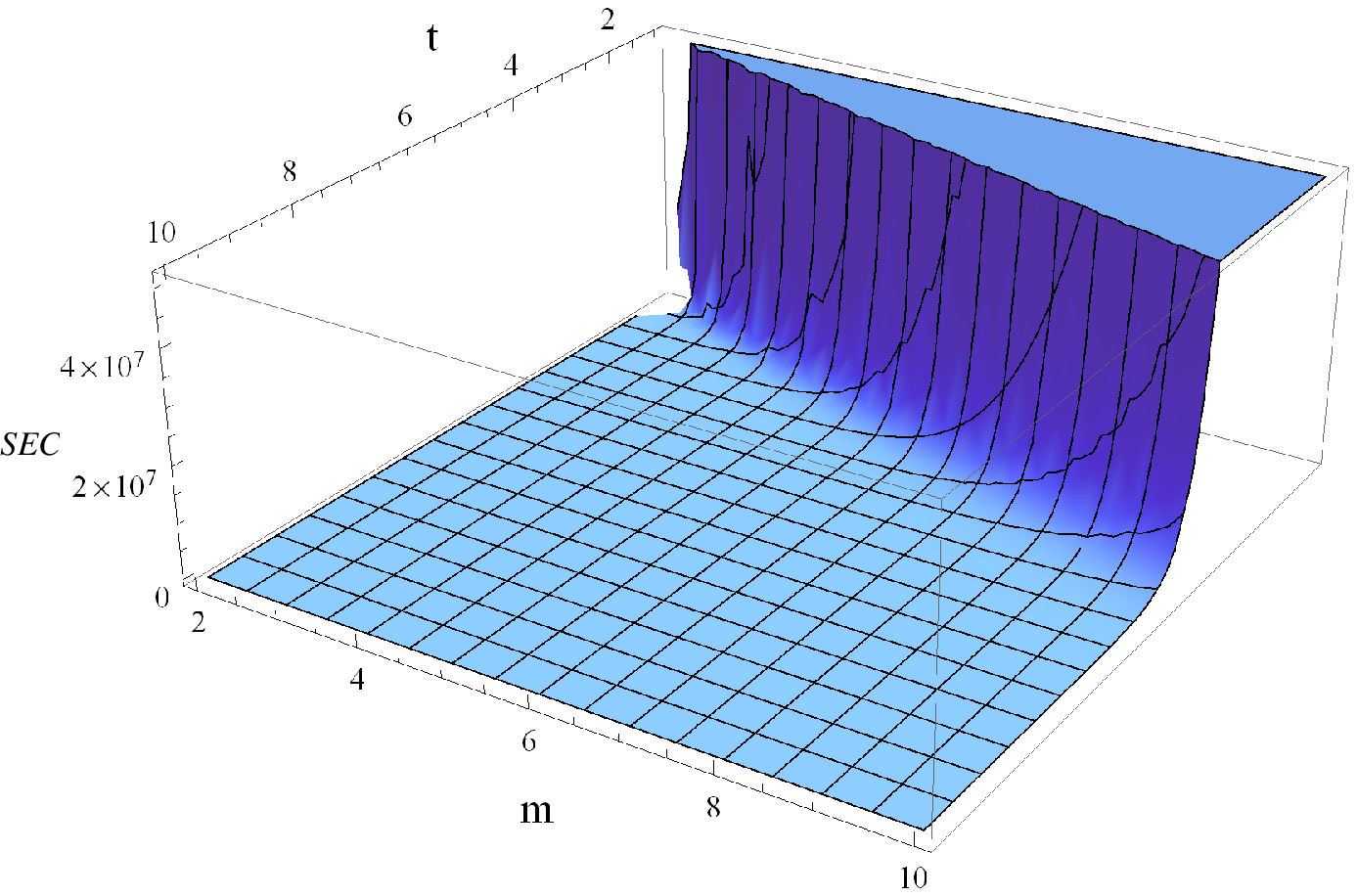, width=0.45\linewidth}
\epsfig{file=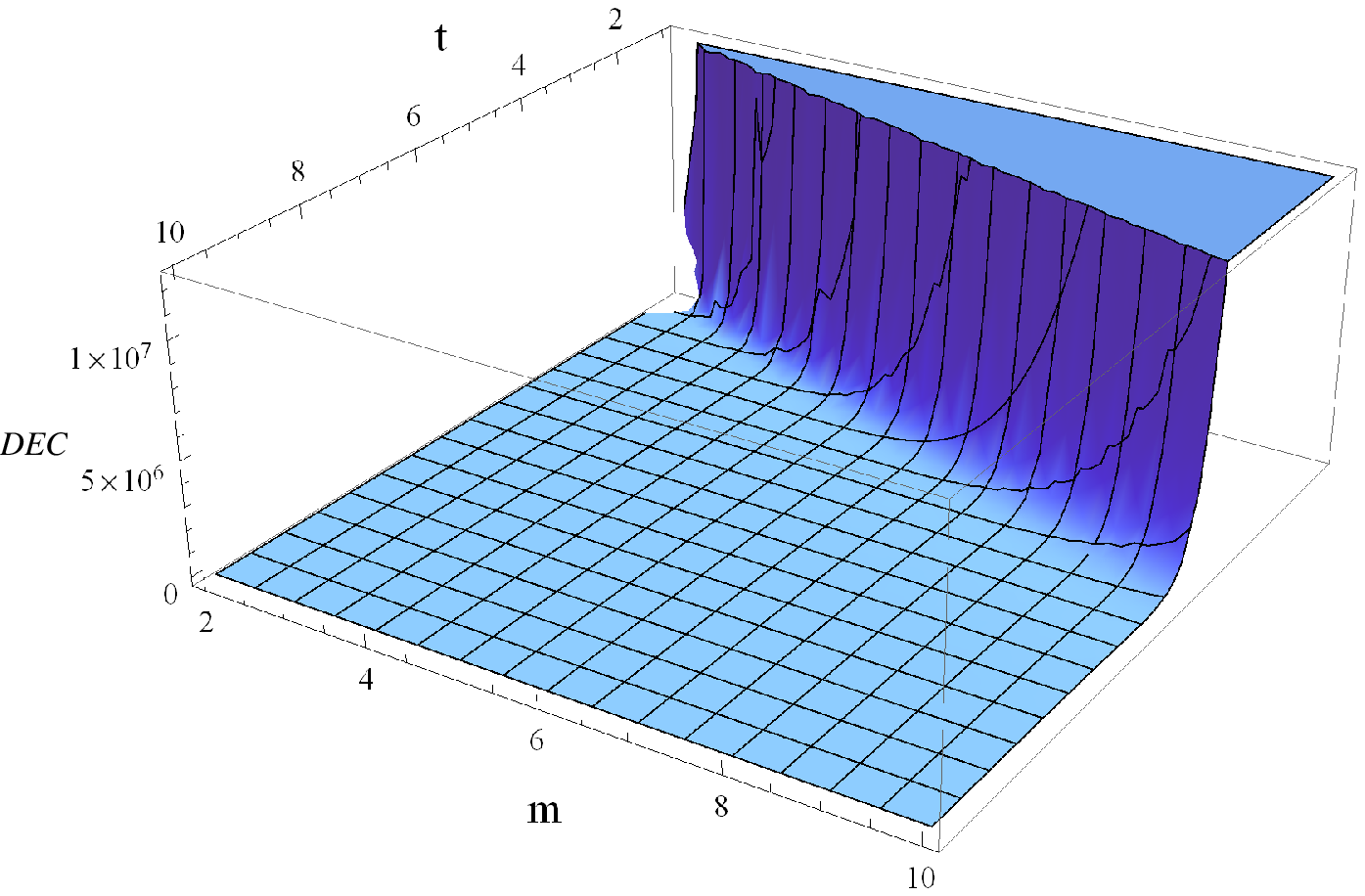, width=0.45\linewidth} \caption{Evolution
of SEC with $\beta_1=0.001$, $\beta_3=0.2$, $\beta_4=0.03$ and DEC
with $\beta_1=0.001$, $\beta_3=-0.2$, $\beta_4=-0.03$ versus $m$ and
$t$.}
\end{figure}

\section{Summary and Discussion}

In this paper, we have formulated the energy constraints in a
general modified theory of gravity involving torsion scalar and a
scalar equivalent to Gauss-Bonnet term. We have taken FRW universe
model filled with perfect fluid matter. Firstly, we have defined
these inequalities for general $F(T,T_G)$ by taking into account the
effective energy density and its pressure. In order to be
particular, we have considered two interesting models of $F(T,T_G)$
recently proposed in literature \cite{24}. We have discussed the
compatibility of the respective energy constraints for these models
by fixing some of the free parameters. In order to examine these
constraints, we have adopted two ways: introduction of some
well-known cosmic parameters like Hubble, jerk, snap and
deceleration parameters (we have used the recent limits of these
parameters that are available in literature) and the power law
cosmology.

Firstly, we explore the compatibility of energy conditions for a
$F(T,T_G)$ model involving four free parameters namely
$\beta_1,~\beta_2,~\alpha_1$ and $\alpha_2$ graphically. In plots,
we have either fixed $\alpha_1,~\alpha_2$ (by assuming their
positive and negative values) and find the possible ranges of
parameters $\beta_1,~\beta_2$ or vice versa. It is seen that WEC can
be satisfied for this model if we take $\alpha_1,~\alpha_2>0$ for
the fixed values fixed $\beta_i$ parameters within the range
$0<\beta_1,~\beta_2<1$. It is interesting to mention here that if we
consider some other positive large values of $\beta_i$, the positive
ranges of $\alpha_i$ still remain valid. Furthermore, if we set
negative values of $\beta_i$ then WEC can be satisfied only for
$\alpha_1,~\alpha_2>0$ if we impose the constraints $\beta_1<0$ and
$\beta_2\leq-2$. In the reverse case where we have fixed parameters
$\alpha_1,~\alpha_2$, either positive small values $0<\alpha_i<1$ or
negative values satisfying $-1<\alpha_i<0$, WEC can be compatible
with this model if $\beta_1,~\beta_2$ satisfies the inequalities
$0\leq\beta_i\leq7$.

Also, NEC constraints can be satisfied for this model if we take
positive ranges of $\alpha_i$ parameters for the specified positive
values of $\beta_i$ parameters. However, such a positive range of
parameters $\alpha_i$ can be achieved so that NEC constraint remains
valid, if one set parameters $\beta_1=-10,~\beta_2=-2$. In a similar
pattern, NEC constraint will be satisfied for small positive and
negative values of $\alpha_i$ parameters with positive large values
of $\beta_i$. In case of SEC, the inequalities will be satisfied
only for $\alpha_i<0$ when we fixed $0<\beta_1,~\beta_2<1$. Further,
if we fix $\alpha_i$ and explore the possible ranges of $\beta_i$ so
that SEC constraint remains compatible with this model, then it is
observed that the inequality holds for $\beta_i>0$ when
$\alpha_i>0$.

In case of power law cosmology, we have fixed all these four
parameters and found the possible variations of parameter $m$ and
the cosmic time. It is seen that for positive fixed values of these
parameters, the inequalities are satisfied $\forall~m,~t>0$ except
the case of SEC where one should fix $\alpha_1=0.001$ for the
purpose.

For the second model involving four parameters
$\beta_1,~\beta_2,~\beta_3$ and $\beta_4$, we have determined the
numerical inequalities given by Eqns.(\ref{48})-(\ref{51}) using the
recent measures of cosmic parameters and discussed them graphically.
It is interesting to mention here that all constraints arising from
WEC, NEC, DEC and SEC are independent of the parameter $\beta_2$. We
have found the possible variations of $\beta_3,~\beta_4$ by fixing
$\beta_1=0.1$ so that the inequalities are satisfied. It is seen
that WEC remains valid if $\beta_3>0$ and $\beta_4>0$. However, for
other constraints corresponding to NEC, SEC and DEC, one have to fix
$\beta_3$ satisfying $\beta_3\gtrless0$. Further, we discuss the
energy constraints using power law solution and found the possible
variations of parameters $m$ and $t$ with all $\beta_i$ fixed. Plots
indicate that in this case, energy constraints can be satisfied for
$\forall m,~t$ using particular values of $\beta_i$. It would be
worthwhile to explore the possible ranges of the involved free
parameters for other models of this gravity by making compatibility
with the energy condition bounds.

\end{document}